\documentclass[12pt,letterpaper]{article}         
\usepackage{jheppub}
\usepackage{epsfig}
\usepackage{amssymb,amsfonts}
\usepackage{color}
\usepackage[usenames,dvipsnames]{xcolor}
\usepackage[T1]{fontenc}
\usepackage[utf8]{inputenc}

\title{\large Chaos and pole-skipping in rotating black holes}

\author{Mike Blake$^{1}$ and Richard A.~Davison$^{2}$}

\affiliation{ $^{1}$ School of Mathematics, Fry Building, Woodland Road, Bristol BS8 1UG, U.K.}

\affiliation{ $^{2}$ Department of Mathematics and Maxwell Institute for Mathematical Sciences, Heriot-Watt University,
Edinburgh EH14 4AS, U.K.}
 
\abstract{We study the connection between many-body quantum chaos and energy dynamics for the holographic theory dual to the Kerr-AdS black hole. In particular, we determine a partial differential equation governing the angular profile of gravitational shock waves that are relevant for the computation of out-of-time ordered correlation functions (OTOCs). Further we show that this shock wave profile is directly related to the behaviour of energy fluctuations in the boundary theory. In particular, we demonstrate using the Teukolsky formalism that at complex frequency $\omega_* = i 2 \pi T$ there exists an extra ingoing solution to the linearised Einstein equations whenever the angular profile of metric perturbations near the horizon satisfies this shock wave equation. As a result, for metric perturbations with such temporal and angular profiles we find that the energy density response of the boundary theory exhibit the signatures of ``pole-skipping'' -- namely, it is undefined, but exhibits a collective mode upon a parametrically small deformation of the profile. Additionally, we provide an explicit computation of the OTOC in the equatorial plane for slowly rotating large black holes, and show that its form can be used to obtain constraints on the dispersion relations of collective modes in the dual CFT.} 

\begin{document}
\maketitle


\section{Introduction}
\paragraph{}The last decade has seen remarkable progress in identifying new characterisations of chaos in many-body quantum systems. In particular, a large amount of attention has focused on the physics of operator scrambling -- namely the increase in size and complexity of operators in a many-body quantum system under Heisenberg time evolution. A valuable tool in probing this time evolution is provided by out-of-time ordered correlation functions (OTOCs), which have been extensively studied in a range of many body systems including quantum circuits \cite{Khemani:2017nda, Nahum:2017yvy}, condensed matter spin systems \cite{Swingle:2016jdj, Kitaev:2017awl, Maldacena:2016hyu} and quantum field theories \cite{Shenker:2013pqa, Roberts:2014isa, Jensen:2016pah, Chowdhury:2017jzb, Turiaci:2016cvo, Shenker:2014cwa}. These developments have led to a range of surprising new insights into many-body quantum chaos, including the discovery of a fundamental bound on the growth of OTOCs in systems with a large on-site Hilbert space dimension \cite{Maldacena:2015waa}, and surprising connections between many-body quantum chaos and hydrodynamics \cite{Grozdanov:2017ajz, Blake:2017ris, Blake:2018leo, Blake:2016wvh, Blake:2021wqj, Blake:2017qgd, Patel:2016wdy, Gu:2016oyy, Davison:2016ngz}.

\paragraph{} Throughout these developments an important role has been played by holographic quantum field theories. In such theories the computation of the OTOC can be reduced to studying a two-particle scattering process near the horizon of a black hole in the dual gravitational description \cite{Shenker:2014cwa}. The spatial and temporal form of the OTOC can then be extracted from the profile of gravitational shock waves at the horizon \cite{Shenker:2014cwa, Roberts:2014isa}. In particular one finds that for theories dual to static, isotropic, planar black holes the initial decay of the OTOC of two local few-body operators $\hat{W}, \hat{V}$ has the form 
\begin{equation}
\langle \hat{W}(t, \vec{x}) \hat{V}(0, 0) \hat{W}(t, \vec{x})  \hat{V}(0, 0) \rangle \sim 1 - c G_N e^{\lambda_L(t - |\vec{x}|/v_B)}
\label{otocintro}
\end{equation}
where $c$ is an order one constant,  $G_N \ll 1$ is Newton's constant in the dual gravitational description, the Lyapunov exponent $\lambda_L = 2 \pi T$ saturates the bound of \cite{Maldacena:2015waa}, and the butterfly velocity $v_B$ is system dependent and characterises the rate of propagation of chaos in the dual quantum field theory. 

\paragraph{} Furthermore, such holographic systems have played an important role in identifying and testing new connections between many-body quantum chaos and hydrodynamics. In particular, at least for maximally chaotic systems, it has been argued that there is remarkable connection between the form of the OTOC~\eqref{otocintro} and an unusual feature of the energy density response of the quantum field theory known as ``pole-skipping'' \cite{Grozdanov:2017ajz, Blake:2017ris}. Namely, for perturbations at specific values of the analytically continued frequency $\omega_* = i \lambda_L$ and wave-numbers $q_* = \pm i \lambda_L/v_B$ related to the form of \eqref{otocintro}, one finds that the energy density response is undefined. Furthermore, the retarded energy-density Green's function exhibits a collective excitation whose dispersion relation $\omega(q)$ passes through the point $(\omega_*, q_*)$.\footnote{In addition to the above example, holographic theories also exhibit pole-skipping in the lower-half plane of complex frequency space \cite{Blake:2019otz}. These are not directly connected to the form \eqref{otocintro} of the OTOC, but can still provide interesting constraints on the dispersion relations of the field theory's collective excitations \cite{Grozdanov:2019uhi,Natsuume:2019xcy,Ceplak:2019ymw,Natsuume:2019vcv,Wu:2019esr,Ahn:2020bks,Grozdanov:2020koi,Ahn:2020baf,Kim:2020url,Natsuume:2020snz,Abbasi:2020xli,Ceplak:2021efc,Kim:2021hqy}.} This ``pole-skipping'' phenomenon arises as a prediction of a hydrodynamic effective theory of maximally chaotic systems \cite{Blake:2017ris, Blake:2021wqj}, and has been shown to be a generic feature of holographic theories dual to planar black holes \cite{Blake:2018leo}. See \cite{Choi:2020tdj} for a proposal for how pole-skipping generalises away from maximal chaos, and \cite{Haehl:2018izb,Jensen:2019cmr,Das:2019tga,Haehl:2019eae,Ramirez:2020qer} for investigations of pole-skipping in CFTs.

\paragraph{} So far, the connection between many-body quantum chaos and pole-skipping in holographic systems has primarily been studied in theories dual to static planar black holes \cite{Blake:2018leo,Grozdanov:2018kkt,Natsuume:2019sfp,Li:2019bgc,Abbasi:2019rhy,Abbasi:2020ykq,Jansen:2020hfd,Sil:2020jhr,Yuan:2020fvv,Jeong:2021zhz}. One important exception is provided by the study of rotating BTZ black holes, which are dual to a 1+1 dimensional CFT on a circle with a chemical potential for rotation.\footnote{See \cite{Ahn:2019rnq,Ageev:2021xjk} for other exceptions.} In such a setting, the form of the OTOC was studied in  \cite{Jahnke:2019gxr, Mezei:2019dfv} (see also \cite{Reynolds:2016pmi,Poojary:2018esz,Halder:2019ric,Banerjee:2019vff,Craps:2020ahu,Craps:2021bmz}), and in \cite{Liu:2020yaf} it has been shown that a natural generalisation of pole-skipping applies for arbitrary values of rotation. Whilst the fact that pole-skipping continues to hold in this case is reassuring, and provides evidence that a hydrodynamic effective theory of chaos continues to apply, both the effects of rotation and the dual gravity description are particularly simple for 1+1 dimensional CFTs.

\paragraph{} In this paper, our goal is to study the relationship between many-body quantum chaos and energy dynamics for an example dual to a higher-dimensional rotating black hole, specifically the Kerr-AdS black hole. Such a black hole, the details of which we review in Section~\ref{sec:kerr}, is dual to a 2+1 dimensional CFT on a sphere with a chemical potential for rotation about the North pole \cite{Hawking:1998kw,Gibbons:2004ai}. In Section~\ref{sec:OTOCsection} we begin by studying the computation of OTOCs in such theories. We note that a previous study of shock waves in the Kerr-AdS black hole has been provided using the Newton-Penrose formalism in \cite{BenTov:2017kyf, BenTov:2019csh}.  Here we provide a complementary description of chaos in which we compute OTOCs using the eikonal approximation to gravitational scattering following \cite{Shenker:2014cwa}. As in previous examples we relate the eikonal phase associated to the gravitational scattering process to the angular profile of a gravitational shock wave at the horizon of the black hole. This angular profile is the Green's function of a second order partial differential equation which we refer to as the shock wave equation, and is given explicitly in equation~\eqref{horizon}. Whilst solving for the angular profile of the shock wave is in general complicated in the Kerr-AdS black hole, an analytically tractable regime is provided by the case of large black holes in the slowly-rotating limit (or limit of small chemical potential). In this regime we determine the spatial profile of the shock wave, and study the form of the OTOC in the equatorial plane.

\paragraph{} We then proceed to discuss the energy density response, and specifically pole-skipping, for the theory dual to the Kerr-AdS black hole. For planar black holes it is well known that the gravitational origin of pole-skipping can be traced to an unusual feature of ingoing solutions to the linearised Einstein equations \cite{Blake:2018leo}. Specifically, when the frequency and wavenumber of the metric perturbations are chosen to take the pole-skipping values $\omega_* = i 2 \pi T$ and  $q_* = \pm i 2 \pi T/v_B$ then one component of the Einstein equations becomes trivial at the horizon, resulting in an extra free parameter in the ingoing solution. Motivated by this, in Section~\ref{sec:HorizonPoleSkippingAnalysis} we analyse the near horizon behaviour of ingoing metric perturbations for the Kerr-AdS background, for which we find an analogous phenomenon. In particular, for metric perturbations at a frequency $\omega_* = i 2 \pi T$ we find that one component of the Einstein equations at the horizon becomes equivalent to the shock wave equation.  As a result, this component of the Einstein equations becomes automatically satisfied for certain choices of the angular profile of the metric near the horizon related to the form of the OTOC.  

\paragraph{} For planar black holes, this observation immediately implies the existence of an extra degree of freedom in the ingoing solution that is the gravitational origin of the pole-skipping phenomenon \cite{Blake:2018leo}.  In order to reach the same conclusion for the Kerr-AdS solution, we use the Teukolsky formalism \cite{Teukolsky:1972my,Teukolsky:1973ha} to provide an analysis of pole-skipping that directly mirrors previous approaches. In particular we use this formalism to confirm the existence of an extra ingoing mode at $\omega_* = i 2 \pi T$ whenever the spatial profile of the near-horizon metric is chosen to satisfy the shock wave equation. Moreover we show that this extra ingoing mode implies the usual signatures of pole-skipping -- namely that the energy density response is undefined, and that a collective mode exists upon a parametrically small deformation of the profile. Finally, for large black holes in the slowly rotating limit we show there is pole-skipping at locations in Fourier space directly related to the temporal and angular profile of the equatorial OTOC, and use such locations to obtain new constraints on the dispersion relations of collective modes of the boundary theory.

\section{The Kerr-AdS Black Hole} 
\label{sec:kerr} 

\paragraph{} As explained in the Introduction, our goal in this paper is to investigate the connection between many-body quantum chaos and energy dynamics in the context of the holographic theory dual to the Kerr-AdS black hole. To begin, it will be useful to review certain features of the Kerr-AdS metric, and to introduce the different coordinate systems that we will use.

\paragraph{} Starting from Boyer-Lindquist coordinates $(t, r, \theta, \phi)$ and then introducing $x = \cos \theta$, the Kerr-AdS metric takes the form 
\begin{equation}
\begin{aligned}
\label{eq:AdSKerrmetric}
ds^2=&\,-\frac{\Delta_r(r)}{\Sigma}\left(dt-\frac{a(1 - x^2)}{\Xi} d\phi\right)^2+\frac{(1-x^2) \Delta_x}{\Sigma} \left( adt-\frac{(r^2+a^2)}{\Xi}d\phi\right)^2\\
&\,+\frac{\Sigma}{\Delta_r(r)}dr^2+\frac{\Sigma}{(1 - x^2) \Delta_x}dx^2,
\end{aligned}
\end{equation}
where the various functions and parameters in \eqref{eq:AdSKerrmetric} are defined as 
\begin{equation}
\begin{aligned}
\label{eq:BGfunctions}
&\Delta_r(r)=(r^2+a^2)\left(1+\frac{r^2}{L^2}\right)-2Mr,\;\;\;\;\;\;\;\;\;\;\;\;\;\Delta_x= 1-\frac{a^2 x^2}{L^2},\\
&\Sigma=r^2+a^2 x^2,\quad\quad\quad\quad\quad\quad\quad\quad\quad\quad\quad\;\;\;\;\;\;\;\;\;\; \Xi=1-\frac{a^2}{L^2},\\
&M=\frac{(r_0^2+a^2)(r_0^2+L^2)}{2L^2r_0}.
\end{aligned}
\end{equation}
The above metric is a solution to Einstein's equations with a negative cosmological constant $\Lambda = - 3/L^2$, with $L$ the AdS-radius. For non-zero $a$ there are two horizons located at the zeroes of $\Delta_r(r)$.  The outer horizon is located at $r=r_0$, in terms of which we may write the temperature $T$ of the black hole as
\begin{equation}
2\pi T\equiv \alpha = \frac{r_0}{2(r_0^2+a^2)}\left(1-\frac{a^2}{r_0^2}+3\frac{r_0^2}{L^2}+\frac{a^2}{L^2}\right).
\label{alpha}
\end{equation}
\paragraph{}The variable $a$ parameterises the rotation of the black hole along the $\phi$ direction. In the coordinate frame $\{t,r,x,\phi\}$, the angular velocity of the outer horizon $\Omega_H$ is given by
\begin{equation}
\Omega_H = \frac{a\Xi}{r_0^2+a^2}.
\end{equation}
Note that $\Omega_H$ is not directly relevant for the thermodynamics of the boundary theory, since the coordinate frame $\{t,r,x,\phi\}$ rotates at infinity with an angular velocity $-a/L^2$. To understand the structure of the boundary theory it is useful to make the coordinate transformation  \cite{Henneaux:1985tv}
\begin{equation}
\begin{aligned}
 R &= \frac{\sqrt{L^2 (r^2 + a^2) -  (L^2 + r^2) a^2 x^2}}{L \sqrt{\Xi}}  ,\quad\quad\quad\quad\Phi = \phi + \frac{a t}{L^2}, \\
\cos \Theta &= \frac{L \sqrt{\Xi} r x }{\sqrt{L^2 (r^2 + a^2) -  (L^2 + r^2) a^2 x^2}},
\end{aligned}
\end{equation}
after which one finds that the asymptotic form of the metric as $R\rightarrow\infty$ is global AdS in standard coordinates
\begin{equation}
\label{eq:AdSKerrmetric2}
ds^2\rightarrow-\bigg(1 + \frac{R^2}{L^2} \bigg) dt^2 + \frac{dR^2}{1 + \frac{R^2}{L^2}} + R^2 (d \Theta^2 + \sin^2 \Theta d \Phi^2).
\end{equation}
The dual field theory then lives on the conformal boundary of global AdS, and corresponds to a 2+1 dimensional CFT on a sphere of radius $L$ with boundary coordinates $\{t, \Theta, \Phi\}$. The temperature of the CFT is given by \eqref{alpha} and there is a chemical potential $\Omega$ for rotation about the North pole given by the angular velocity of the horizon relative to the conformal boundary \cite{Caldarelli:1999xj,Hawking:1998kw,Gibbons:2004ai}
\begin{equation}
\Omega = \frac{a}{r_0^2 + a^2} \bigg(1 + \frac{r_0^2}{L^2} \bigg).
\end{equation}
In other words, the density matrix of the boundary theory is
\begin{equation}
\label{densitymatrix} 
\rho =  \frac{e^{- \beta H - \beta\Omega J }}{Z},\quad\quad\quad\quad\quad\quad\quad\quad   Z = \mathrm{Tr}(e^{- \beta H - \beta\Omega J}),
\end{equation}
where $\beta = 1/T$ and $J$ is the angular momentum about the North pole. 

\paragraph{} In order to study OTOCs, we also need to introduce Kruskal-like coordinates in which the metric is manifestly smooth at the outer horizon. To do this, following \cite{BenTov:2019csh}, we first introduce the co-rotating angular coordinate
\begin{equation}
\label{eq:psidefn}
\quad\quad \psi=\phi - \Omega_H t. 
\end{equation} 
The singularities in the metric at the outer horizon can then be removed by introducing ``ingoing'' ($v$) and ``outgoing'' ($u$) co-ordinates defined by
\begin{equation}
\label{eq:nullcoordsdefn}
du\equiv dt - \frac{(r^2+a^2)}{\Delta_r}dr,\quad\quad\quad\quad\quad dv=dt+\frac{(r^2+a^2)}{\Delta_r}dr, 
\end{equation}
from which Kruskal-like co-ordinates can be constructed as
\begin{equation}
\label{eq:UVcoorddefn}
U=-e^{-\alpha u},\quad\quad\quad\quad\quad\quad\quad\quad V=e^{\alpha v}.
\end{equation}
\paragraph{} In these co-ordinates, the metric can be written as 
\begin{equation}
\begin{aligned}
ds^2=&-\frac{\Delta_r}{\Sigma}\left[\frac{r_0^2+a^2 x^2}{2\alpha(r_0^2+a^2)}\left(\frac{dV}{V}-\frac{dU}{U}\right)-\frac{a(1-x^2)}{\Xi}d\psi\right]^2+\frac{\Sigma\Delta_r}{4\alpha^2(r^2+a^2)^2}\left(\frac{dV}{V}+\frac{dU}{U}\right)^2\\
&+\frac{(1-x^2)\Delta_x}{\Sigma}\left[\frac{a(r_0^2-r^2)}{2\alpha(r_0^2+a^2)}\left(\frac{dV}{V}-\frac{dU}{U}\right)-\frac{(r^2+a^2)}{\Xi}d\psi\right]^2+\frac{\Sigma}{(1 - x^2) \Delta_x}dx^2,
\label{kruskal}
\end{aligned}
\end{equation}
where, as usual, factors of $r$ should be understood as meaning $r(U V)$ as determined through equations \eqref{eq:nullcoordsdefn} and \eqref{eq:UVcoorddefn}. The metric \eqref{kruskal} is smooth at $r = r_0$, corresponding to $U V = 0$, and explicit expressions for the individual metric components in these coordinates can be found in Appendix~\ref{app:kruskalcoords}.

\section{OTOCs in Kerr-AdS black holes} 
\label{sec:OTOCsection}

\paragraph{} We will now discuss the computation of the out-of-time ordered correlation functions for the theory dual to \eqref{eq:AdSKerrmetric}. As is now well known, for holographic systems dual to classical gravity there is a well understood prescription to compute the OTOC of the dual field theory in terms of high-energy scattering process near the horizon of a black hole \cite{Shenker:2014cwa}.  So far, however, computations of the OTOC in theories dual to rotating black holes have primarily been restricted to the case of the BTZ black hole \cite{Jahnke:2019gxr, Mezei:2019dfv} (see also \cite{Reynolds:2016pmi,Poojary:2018esz,Halder:2019ric,Banerjee:2019vff,Craps:2020ahu,Craps:2021bmz}). We note that a previous discussion of shock waves in Kerr-AdS black holes in the context of the Newton-Penrose formalism has been provided by \cite{BenTov:2019csh, BenTov:2017kyf}. However the connection between the formalism in \cite{BenTov:2019csh, BenTov:2017kyf} and the usual computation of OTOCs in terms of gravitational shock waves is non-trivial. Here we will provide a complimentary discussion of the computation of OTOCs in Kerr-AdS black holes in which we work directly with the metric variables following \cite{Shenker:2014cwa}. In particular we will derive the partial differential equation that governs the angular profile of gravitational shock waves, and use this to explicitly analyse the form of the equatorial OTOC for large black holes in the slowly rotating limit.  

\paragraph{} For concreteness, we will be interested in computing correlation functions in the boundary CFT dual to \eqref{eq:AdSKerrmetric} of the form\footnote{We assume that the OTOC in \eqref{otoc} has been regulated by a small separation of operators in imaginary time.}%
\begin{equation}
\langle \hat{W}(t_2, \Theta_2, \Phi_2) \hat{V}(t_1, \Theta_1, \Phi_1) \hat{W}(t_2, \Theta_2, \Phi_2) \hat{V}(t_1, \Theta_1, \Phi_1) \rangle,
\label{otoc} 
\end{equation}
where  $t_2 - t_1 \gg \beta$ and the expectation value is taken with respect to the density matrix \eqref{densitymatrix}. Whilst the OTOC in \eqref{otoc} is a one-sided quantity, it is in practice most efficiently computed using the Kruskal geometry \eqref{kruskal}, dual to a thermofield double state.  The OTOC \eqref{otoc} can be related in such a geometry to a two particle gravitational scattering process between particles moving along the $U=0$ and $V=0$ horizons \cite{Shenker:2014cwa} .

\paragraph{} In particular, in a symmetric frame, perturbing the thermofield double state dual to \eqref{kruskal} by a local boundary operator $\hat{W}$ at time $t_2$ will result in quanta moving along the $V=0$ horizon with momentum $p_1^{U}$.\footnote{Note we are interested in the one-sided OTOC \eqref{otoc} and hence all operators act on the right hand boundary.}  Likewise a perturbation by an operator $\hat{V}$ at time $t_1$ results in quanta moving along the $U=0$ horizon with momentum $p_2^{V}$, highly boosted relative to the $\hat{W}$ quanta such that $p_1^U p_2^V \sim e^{2 \pi T (t_2 - t_1)}$. At the level of the classical gravity approximation the scattering process is elastic, and conserves both the momenta and angular coordinates of the quanta at the horizon \cite{Shenker:2014cwa}. As such the scattering amplitude takes the form $e^{i \delta}$, where the eikonal phase $\delta$ is real and a function of $p_1^{U}, p_2^{V}$ and the horizon coordinates of the quanta. The OTOC \eqref{otoc} is then given by an integral of the scattering amplitude $e^{i \delta}$ over momenta and horizon coordinates, weighted by bulk-to-boundary wave functions that decompose the perturbations caused by the boundary operators into horizon quanta \cite{Shenker:2014cwa}.

\paragraph{} Our goal in this section is to compute the eikonal phase shift  $\delta$ for gravitational scattering in the Kerr-AdS black hole~\eqref{kruskal}. To this end we consider particles whose trajectories approach the null geodesic given by $(U=U(\tau), V(\tau) = 0, x(\tau) = x_1, \psi(\tau) = \psi_1)$, corresponding to quanta created by the $\hat{W}$ operator. Note that the fact we are in co-rotating coordinates is essential to ensure such a trajectory obeys the geodesic equations near the horizon. The stress tensor of such a particle is then given by~\cite{Shenker:2014cwa, Balasubramanian:2019stt}
\begin{eqnarray}
T^{UU} &=& \frac{1}{\sqrt{-g}} p_1^{U} \delta(V) \delta(x - x_1) \delta(\psi - \psi_1), \nonumber \\
 T^{\mu \nu} &=& 0, \;\;\;\; \mu,\nu \neq U.
 \label{source1}
\end{eqnarray}
Additionally we consider a second particle whose trajectory approaches the null geodesic $(V = V(\tau), U(\tau) = 0, x(\tau) = x_2, \psi(\tau) = \psi_2)$, corresponding to quanta created by the $\hat{V}$ operator. This has a stress-energy tensor:
\begin{eqnarray}
T^{VV} &=& \frac{1}{\sqrt{-g}} p_2^{V} \delta(U) \delta(x - x_2) \delta(\psi - \psi_2), \nonumber \\
 T^{\mu \nu} &=& 0, \;\;\;\; \mu,\nu \neq V.
 \label{source2}
\end{eqnarray}
 \paragraph{}   The eikonal phase for gravitational scattering can then be computed by determining the gravitational backreaction $\delta g_{\mu \nu}$ sourced by these particles. The eikonal phase is then given by evaluating the on-shell action to linear order in $\delta g_{\mu \nu}$, which can be written as~\cite{Shenker:2014cwa, Balasubramanian:2019stt}\footnote{Here we ignore non-linearities corresponding to interactions between the two shock-waves, which are higher order in $G_N$.} %
\begin{equation}
\delta = S_{EH}[\delta g_{\mu \nu}] = \frac{1}{4} \int d^4x \sqrt{-g} \delta g_{\mu \nu} T^{\mu \nu} = \frac{1}{4} \int d^4 x \sqrt{-g} \bigg(\delta g_{UU} T^{UU} + \delta g_{VV} T^{VV}\bigg).
\label{onshell} 
\end{equation}
 \subsection*{Gravitational backreaction} 
\paragraph{} We now wish to compute the gravitational backreaction due to the stress tensors \eqref{source1} and \eqref{source2}, which as usual will take the form of gravitational shock waves across the $U=0$ and $V=0$ horizons. Let us begin by determining the backreaction of the particle moving along the $V=0$ horizon, with stress tensor \eqref{source1}. After lowering indices the only non-zero component of the stress tensor $T_{\mu \nu}$ is the $VV$ component, which we can explicitly write as 
\begin{equation}
T_{VV} = \frac{ \Delta_r'(0)  K_0 \Sigma_0(x)}{2 \alpha^2 (r_0^2 + a^2)^2} p_1^{U} \delta(V) \delta(x - x_1) \delta(\psi - \psi_1) ,
\label{stresstensor1}
\end{equation}
where 
\begin{equation}
K_0=\frac{\Xi}{r_0^2+a^2},\quad\quad \Sigma_0(x)=r_0^2+a^2x^2,  \quad\quad \Delta_r'(0) = \frac{d\Delta_r(UV)}{d(UV)}\bigg|_{UV = 0}.
\end{equation}
\paragraph{} To solve for the backreaction, we need to find a metric that satisfies the Einstein equations with the stress tensor \eqref{stresstensor1}. Following~\cite{Shenker:2014cwa} we take an ansatz corresponding to a shock wave metric in which we consider two sides of the Kruskal geometry \eqref{kruskal} with a shift $U \to U + f_1(x, \psi)$ as we cross the $V=0$ horizon.  As we show in Appendix~\ref{app:kruskalcoords}, to linear order in $f_1(x,  \psi)$ such a shock wave can be described by a metric perturbation of the form 
\begin{equation}
\delta ds^2 = - 2 g_{UV}(0) f_1(x, \psi) \delta(V) dV^2  + \dots,   \;\;\;\;\; g_{U V}(0) = \frac{\Delta_r'(0)  \Sigma_0(x)}{2 \alpha^2(r_0^2 + a^2)^2}, 
\label{ansatz}
\end{equation}
where $\dots$ indicate terms proportional to $V^{N} \delta(V)$ where $N \geq 1$ is an integer. Such terms are formally zero and so we ignore them.\footnote{The shock wave metric ansatz in \cite{BenTov:2019csh, BenTov:2017kyf} includes such terms. We have verified that this ansatz, including the $V^N\delta(V)$ terms, leads to the same shock wave equation we obtain below.}

\paragraph{} Inserting the ansatz \eqref{ansatz} into the Einstein equations, with stress tensor given by \eqref{stresstensor1}, we find the only non-trivial equation is given by evaluating the $VV$ component of the Einstein equation
\begin{equation}
R_{VV}-\frac{1}{2}Rg_{VV}-\frac{3}{L^2}g_{VV}= 8 \pi G_N T_{VV},
\end{equation}
at the $V=0$ horizon. We then find the ansatz \eqref{ansatz} obeys the sourced Einstein equations provided the angular profile $f_1(x,\psi)$ obeys the differential equation
\begin{equation}
\label{horizon}
{\cal L} f_1 = 8 \pi G_N K_0 \Sigma_0(x) p_1^{U} \delta(x - x_1) \delta(\psi - \psi_1), 
\end{equation}
where ${\cal L}$ is the second order differential operator
\begin{equation}
\begin{aligned}
\mathcal{L}=&\,\partial_x(\Delta_x(x)(1-x^2)\partial_x)+\frac{K_0^2\Sigma_0(x)^2}{(1-x^2)\Delta_x(x)}\left(\partial_\psi^2 +\frac{2a(1-x^2)}{K_0\Sigma_0(x)}\left(\alpha+\frac{r_0\Delta_x(x)}{\Sigma_0(x)}\right)\partial_\psi \right)\\
&\,-\alpha\left(2r_0-\frac{a^2\alpha(1-x^2)}{\Delta_x(x)}\right). 
\end{aligned}
\label{horizondiff}
\end{equation}
Equations \eqref{horizon} and \eqref{horizondiff} are the central result of this section. The angular profile of a gravitational shock wave on the $V=0$ horizon is uniquely determined by solving \eqref{horizon} subject to periodicity under $\psi \to \psi + 2 \pi$ and regularity at $x=\pm 1$. 

\paragraph{}There is an entirely analogous discussion for the gravitational backreaction of the second particle \eqref{source2} moving on the $U=0$ horizon. The backreaction of this particle can be described by a shock wave metric arising from a shift $V \to V + f_2(x,\psi)$ across the $U=0$ horizon. To linear order in $f_2$ this corresponds to the metric perturbation
\begin{equation}
\delta ds^2 = -2 g_{UV}(0) \delta (U) f_2(x,\psi)dU^2,
\label{ansatz2}
\end{equation}
which satisfies the Einstein equations with source \eqref{source2} provided $f_2(x, \psi)$ satisfies
\begin{equation}
\begin{aligned}
\tilde{\cal L} f_2 = 8 \pi G_N K_0 \Sigma_0(x) p_2^{V}   \delta(x - x_2) \delta(\psi - \psi_2) ,
\end{aligned}
\end{equation}
where $\tilde{\cal L}$ is the differential operator appearing in \eqref{horizondiff} but with $\partial_{\psi} \to -{\partial}_{\psi}$.

\paragraph{} Finally, to determine the eikonal phase using \eqref{onshell}, it is helpful to first define $f_0(x, x', \psi)$ as the periodic, regular solution to the equation 
 \begin{equation}
 {\cal L} f_0 = -\delta(x - x') \delta(\psi).
 \label{eq:normshock}
 \end{equation}
The shock wave solutions \eqref{ansatz} and \eqref{ansatz2} then correspond to a metric whose non-zero components take the form $\delta g_{VV} \sim f_0(x, x_1, \psi - \psi_1)$ and $\delta g_{UU} \sim f_0(x,x_2,\psi_2-\psi)$. Using the symmetry of the Green's function $f_0(x_1,x_2, \psi) = f_0(x_2,x_1, \psi)$, we then find the on-shell action \eqref{onshell} evaluates to
\begin{equation}
\delta(p_1^{U}, p_2^{V}, x_1,x_2, \psi_2 - \psi_1) =   4 \pi G_N \frac{K_0 \Delta_r'(0) \Sigma_0(x_1) \Sigma_0(x_2)}{\alpha^2 (r_0^2 + a^2)^2} p_1^{U} p_2^{V} f_0(x_2, x_1, \psi_2 - \psi_1) . 
\label{phase}
\end{equation}
\paragraph{} The expression \eqref{phase} for the eikonal phase shift fully characterises the gravitational scattering process relevant for many-body quantum chaos. Formally the expression \eqref{phase} is similar to that for planar black holes presented in \cite{Shenker:2014cwa} -- for the relevant values of momenta $p_1^{U}, p_2^{V}$ then in co-rotating coordinates the eikonal phase grows exponentially as $e^{2 \pi T (t_2 - t_1)}$, and has an angular profile related to the shock wave profile $f_0$. However using \eqref{phase} to explicitly compute the form of the OTOC \eqref{otoc} is significantly more complicated than for BTZ or planar black holes, in particular due to the difficulty in solving the PDE \eqref{horizon} directly.  As such we will now restrict to studying the form of the equatorial OTOC for slowly rotating, large black holes, which we will be able to determine analytically.

\subsection{Equatorial OTOC in Kerr-AdS black holes} 
\label{sec:slowrototoc}

 \paragraph{} We now wish to consider the form of the OTOC \eqref{otoc} in the equatorial plane, that is we set $\Theta_1 = \Theta_2 = \pi/2$ and consider the correlation function
 \begin{eqnarray}
 \label{OTOC} 
 H_4(t, \Phi) \equiv
 \langle \hat{W}(t_2, 0, \Phi_2) \hat{V}(t_1, 0, \Phi_1) \hat{W}(t_2, 0, \Phi_2) \hat{V}(t_1, 0, \Phi_1)  \rangle,
 \end{eqnarray}
 %
 where $t = t_2 - t_1$ and $\Phi = \Phi_2 - \Phi_1$. For such operator insertions then, for large black holes $r_0/L \gg 1$, the angular profiles of the bulk-to-boundary wavefunctions relevant for the OTOC will be sharply peaked at the horizon around the equatorial plane $x_1 \approx x_2 \approx 0 $ and around co-rotating angles $\psi_{1,2}$ such that $\psi_2 - \psi_1 \approx \Phi - \Omega t$. To leading order in $G_N$, the exponential growth of the equatorial OTOC is then proportional to the eikonal phase \eqref{phase} evaluated at the relevant coordinates and momenta for the scattering process (see e.g. \cite{Shenker:2014cwa, Blake:2021wqj}). This gives 
 \begin{eqnarray}
 \label{OTOC2} 
 H_4(t, \Phi) \sim 1 - c G_N e^{2\pi T t} f_0(0, 0, \Phi - \Omega t ), 
 \end{eqnarray}
where $c$ is an order one constant. In general the shock wave equation obeyed by $f_0(x, 0, \psi)$ is still a complicated PDE, and we will not attempt to solve it generically in this paper. However, in the slowly rotating limit $a \ll r_0$, $a \ll L$ it is possible to determine $f_0(x, 0, \psi)$ analytically, and hence determine the form of the equatorial OTOC \eqref{OTOC2} for slowly rotating, large black holes.  

\paragraph{}In particular, in the slowly rotating limit $a \ll r_0$, $a \ll L$ then equation~\eqref{eq:normshock} for the normalised shock wave reduces, for a source in the equatorial plane, to
\begin{equation}
\left(\partial_x((1-x^2) \partial_x)+\frac{\partial^2_{\psi}}{1-x^2} +3\frac{a}{r_0}\left(1+\frac{r_0^2}{L^2}\right) \partial_{\psi} - \left(1+\frac{3r_0^2}{L^2}\right)  \right)f_0(x, 0, \psi)= -\delta(x) \delta(\psi),
\label{shockslowly}
\end{equation}
which we need to solve subject to periodicity under $\psi \to \psi + 2 \pi$ and regularity at the North and South poles (i.e.~$x = \pm 1$). As we discuss in Appendix~\ref{app:slowlyrotating}, an exact solution to the homogeneous version of the shock wave equation \eqref{shockslowly} can be obtained in terms of Legendre functions, yielding the following Fourier series representation for $f_0(0,0, \psi)$ (up to an overall normalisation) 
\begin{equation}
f_0(0, 0, \psi)= \sum_{k\in\mathbb{Z}} \frac{\Gamma\left[\frac{1}{4}-\frac{k}{2}-\frac{i\nu}{2} \right] \Gamma\left[\frac{1}{4}-\frac{k}{2}+\frac{i\nu}{2} \right]}{\Gamma\left[\frac{3}{4}-\frac{k}{2}-\frac{i\nu}{2} \right]\Gamma\left[\frac{3}{4}-\frac{k}{2}+\frac{i\nu}{2} \right]}  e^{i k \psi},
\label{otocrotating} 
\end{equation}
where
\begin{equation}
\label{eq:nudefn}
\nu=\frac{\sqrt{3}}{2}\sqrt{1+\frac{4r_0^2}{L^2}-\frac{4ika}{r_0}\left(1+\frac{r_0^2}{L^2}\right)}. 
\end{equation}

\paragraph{}We now take the large black hole limit by taking the limit $r_0/L \gg 1$ in \eqref{otocrotating} whilst holding $L \psi/r_0 \sim 1$ fixed. As explained in Appendix~\ref{app:slowlyrotating}, from this we find that for $\frac{L}{r_0}\ll |\psi|\ll 1$ the shock wave profile is given by 
\begin{equation}
\label{otoccorotating}
f_0(0, 0, \psi)\propto\frac{1}{\sqrt{|\nu_0\psi|}}\begin{cases} \exp\left(-\left(1+\frac{\sqrt{3}a}{2L}\right)|\nu_0\psi|\right) & \psi>0,\\ \exp\left(-\left(1-\frac{\sqrt{3}a}{2L}\right)|\nu_0\psi|\right) & \psi<0,\end{cases}
\end{equation}
where $\nu_0 = \sqrt{3} r_0/L$. %

\paragraph{} From this shock wave profile it is straightforward to extract the equatorial OTOC using \eqref{OTOC2}. We find that for $\frac{L}{r_0} \ll |\Phi - \Omega t| \ll 1$ the equatorial OTOC of slowly rotating, large black holes can be written in the form 
\begin{equation}
H _4(t, \Phi) \propto\frac{G_N}{\sqrt{|\nu_0 (\Phi - \Omega t)|}}\begin{cases} \exp\left(2 \pi T_+ (t - \frac{\Phi L}{v_B^+} )\right) & \Phi - \Omega t >0,\\ \exp\left(2 \pi T_- (t  + \frac{\Phi L}{v_B^-} )\right) & \Phi - \Omega t < 0, \end{cases}
\label{equatorialOTOC} 
\end{equation}
where
\begin{equation}
\label{lyapunov}
2 \pi T_\pm = 2 \pi T \bigg(1 \pm \frac{2 a}{\sqrt{3} L} + \dots \bigg) = 2 \pi T \left( 1 \pm \frac{L \Omega}{v_B^{(0)}} + \dots \right), 
\end{equation}
 and
  \begin{equation}
 \label{exponents}
 \frac{2 \pi T_\pm}{v_B^\pm}   =  \frac{\nu_0}{L}\bigg(1 \pm \frac{\sqrt{3} a}{2 L} + \dots \bigg) = \frac{2 \pi T}{v_B^{(0)}}\bigg(1 \pm  L \Omega v_B^{(0)} + \dots \bigg).
 \end{equation} 
 Note that to this order in the large black hole and slowly rotating limit we have $2 \pi T = 3 r_0/2L^2$, $\Omega = a/L^2$ and $v_B^{(0)} = \sqrt{3}/2$ is the butterfly velocity of the  non-rotating black hole.\footnote{Care must taken in interpreting the exponents $2 \pi T_\pm$, since the functional form \eqref{equatorialOTOC} is only valid for the regime $|\Phi - \Omega t | \ll 1$. In particular, as for the rotating BTZ black hole \cite{Mezei:2019dfv}, the functional form \eqref{equatorialOTOC} does not imply a violation of the chaos bound \cite{Maldacena:2015waa}.} 
 
 \paragraph{}The structure of the corrections in equations \eqref{lyapunov} and \eqref{exponents} can be understood by noting that the form \eqref{equatorialOTOC} is equivalent to that obtained by boosting the OTOC of a non-rotating large black hole $e^{2 \pi T(t - |\Phi|L/v_B^{(0)})}/\sqrt{|\Phi|}$ by a velocity $v = - \Omega L \ll 1$. We also note that for this reason, when expressed in terms of the variables $\{\Omega, T, L, v_B^{(0)}\}$, the corrections to the OTOC of a decompactified BTZ black hole presented in~\cite{Mezei:2019dfv} can be written in the same form as \eqref{lyapunov} and \eqref{exponents} for $\Omega L \ll 1$. An important physical difference to the case of BTZ however is found in the corrections to the equatorial butterfly velocities $v_B^{\pm}$. In particular it was observed in ~\cite{Mezei:2019dfv} that the butterfly velocity of rotating BTZ black holes is independent of the chemical potential for rotation. However, from \eqref{lyapunov} and \eqref{exponents} we find that the equatorial butterfly velocities $v_B^{\pm}$ of large Kerr-AdS black holes are corrected at leading order in $\Omega L$ to 
\begin{equation}
v_B^{\pm} = \frac{\sqrt{3}}{2} \bigg( 1 \pm \frac{1}{2 \sqrt{3} } \frac{a}{L} + \dots \bigg) = v_B^{(0)}\left(1 \pm \left(\frac{1}{v_B^{(0)}} - v_B^{(0)} \right) L \Omega + \dots \right).
\label{butterfly}
\end{equation}
In this context, the fact that there is no correction to the butterfly velocity in the BTZ case is related to the fact that $v_B^{(0)} = 1$ for such a system. 

\section{Near-horizon metric perturbations of Kerr-AdS}
\label{sec:HorizonPoleSkippingAnalysis}

\paragraph{} In the last Section we studied the computation of the OTOC in the CFT dual to the Kerr-AdS black hole, and in particular we showed that the angular profile of the gravitational shock waves relevant for the OTOC are governed by the shock wave equation~\eqref{horizon}. For the case of large black holes in the slowly rotating limit we were also able to obtain an explicit form for the equatorial OTOC as given by~\eqref{equatorialOTOC}. We now turn to examine the connection between these features of the OTOC and the linear response properties of the field theory (in particular its energy density response), which are encoded in ingoing metric perturbations around the geometry \eqref{eq:AdSKerrmetric}.

\paragraph{}In particular, as we reviewed in the Introduction, for the simpler case of planar AdS black holes it is known that at certain carefully chosen (complex) values of frequency $\omega$ and wavenumber $q$ there surprisingly exist extra ingoing solutions to the linearised Einstein's equations \cite{Grozdanov:2017ajz,Blake:2018leo,Blake:2019otz}. One implication of this fact is that at these carefully chosen values these spacetimes necessarily support a quasinormal mode, corresponding to a collective mode of the boundary theory. Furthermore, the retarded Green's function of the energy density operator in the dual field theory exhibits a ``pole-skipping'' phenomenon, characterised by a non-uniqueness of the (complex) Fourier space Green's function due to the intersection of a pole and a zero.

\paragraph{}Although explicitly solving the relevant perturbation equations is difficult, the existence of these extra ingoing solutions can easily be inferred just from studying the perturbation equations near the horizon in ingoing coordinates. One particularly interesting instance can be inferred by examining the $vv$ component of the Einstein equations (where $v$ is the ingoing coordinate) on the horizon \cite{Blake:2018leo}. This equation vanishes identically for modes of a specific frequency and wavenumber $(\omega_*, q_*)$, indicating that there is an extra free parameter in the general ingoing solution. This instance is particularly interesting as in general $\omega_*= i 2 \pi T$ and $q_* = \pm i 2 \pi T/v_B$, where  $2\pi T$ and $v_B$ are the Lyapunov exponent and butterfly velocity governing the temporal and spatial profile of the OTOC. This universal result provides evidence for a hydrodynamic effective theory of chaos in such systems \cite{Blake:2017ris}. 

\paragraph{}In anticipation of the fact that the Kerr-AdS black hole does not have spatial translational invariance, it will be helpful to rephrase the origin of the pole-skipping phenomenon described above in a way that does not rely on a spatial Fourier transform. In particular, if one studies Fourier modes of frequency $\omega_* = i 2 \pi T$ in the above-mentioned planar black holes, then the $vv$ component of the Einstein equations on the horizon becomes a second order differential equation for the spatial profile of $\delta g_{vv}$ at the horizon. This differential equation is identical to the one governing the spatial profile of a gravitational shock wave on the $V=0$ horizon, in the absence of the delta function source. Therefore if the spatial profile of $\delta g_{vv}$ at the horizon is chosen to satisfy the (sourceless) shock wave equation then the $vv$ component of the Einstein equation is automatically satisfied, and it can be shown this results in an extra ingoing solution to the linearised Einstein equations \cite{Blake:2018leo}. 
 
\paragraph{}In the remainder of this paper we will explain if and how such phenomena generalise to the case of the Kerr-AdS black hole. In this Section we will focus only on the near-horizon properties of metric perturbations, showing that at $\omega_* = i 2 \pi T$\footnote{Throughout this paper $\omega$ will refer to the frequency of perturbations in co-rotating coordinates.} the $vv$ component of the Einstein equations again reduces to a differential equation for the angular profile of $\delta g_{vv}$ on the horizon, which is equivalent to that implied by the sourceless shock wave equation \eqref{horizon}. We will deal carefully with the radial evolution to the boundary, and the connection to field theory linear response, in Section \ref{sec:QNMSection}.

\paragraph{}Since we are interested in ingoing metric perturbations about the spacetime \eqref{eq:AdSKerrmetric}, we work in coordinates $(v, r, x,\psi)$ where $v$ is the ingoing coordinate defined in equation \eqref{eq:nullcoordsdefn}, and $\psi$ is the co-rotating angular coordinate defined in \eqref{eq:psidefn}. We are then interested in studying linear perturbations of the metric, after Fourier transforming with respect to the ingoing coordinate $v$
\begin{equation} 
\label{metpert}
\delta g_{\mu \nu}(v, r, x, \psi)  = e^{-i \omega v}\delta g_{\mu \nu}(r, x, \psi),
\end{equation} 
and, as motivated above, we wish to examine the form of the Einstein equations for $\omega = \omega_* = i 2 \pi T$. The modes \eqref{metpert} obey a complicated set of coupled partial differential equations. However, motivated by the results for planar black holes described above, for now we will focus on just the $vv$-component of the Einstein equations at the outer horizon $r=r_0$. Assuming that all of the metric perturbations are regular at the horizon,\footnote{Specifically that the metric perturbations and their first and second derivatives are finite at $r=r_0$.} then for $\omega = i 2 \pi T$ evaluating this equation at the horizon gives
\begin{equation}
\begin{aligned}
\label{eq:NHgvveq}
&\partial_x\left((1-x^2)\Delta_x(x) \partial_x\frac{\delta g_{vv}(r_0,x,\psi)}{\Sigma_0(x)}\right) -\alpha \left(2r_0-\frac{a^2 \alpha(1-x^2)}{\Delta_x(x)}\right)\frac{\delta g_{vv}(r_0,x,\psi)}{\Sigma_0(x)} \\
+&\frac{K_0^2 \Sigma_0(x)^2}{(1-x^2)\Delta_x(x)}\Biggl[\partial_{\psi}^2\frac{\delta g_{vv}(r_0,x,\psi)}{\Sigma_0(x)} +\frac{2a (1-x^2)}{K_0 \Sigma_0(x)}\left(\alpha+\frac{r_0\Delta_x(x)}{\Sigma_0(x)}\right)\partial_{\psi}\frac{\delta g_{vv}(r_0,x,\psi)}{\Sigma_0(x)}\Biggr] =0.
\end{aligned}
\end{equation}
\paragraph{}Crucially, this horizon equation depends only on $\delta g_{vv}(r_0,x,\psi)$ ands its angular derivatives at the horizon -- all other metric components have decoupled from this equation.  Moreover, from comparison to \eqref{eq:NHgvveq} we see that this horizon equation is nothing but the sourceless version of the shock wave equation \eqref{horizon} that controls the spatial profile of the OTOC, where $\delta g_{vv}(r_0, x, \psi)/\Sigma_0(x)$ appears in the role of the shift $f_1$ as is natural from equation~\eqref{ansatz}.

\paragraph{}Now, any ingoing solution to the linearised Einstein equations must necessarily satisfy the horizon equation \eqref{eq:NHgvveq}. Hence either the metric component $\delta g_{vv}$ must vanish everywhere on the horizon, or it must have a specific spatial profile there, corresponding to a non-trivial solution of the equation \eqref{eq:NHgvveq}. The key point is the existence of the latter option -- it indicates that if $\delta g_{vv}(r_0, x, \psi)$ has an appropriate spatial profile then it is possible for it to be non-zero on the horizon, allowing for the potential existence of an extra ingoing solution.

\paragraph{}This argument can be made precise in the context of planar black holes: for modes of an appropriate spatial profile there is an additional parameter in the ingoing solution to the linearised Einstein's equations and this is the origin of pole-skipping \cite{Blake:2018leo}. For the case of the Kerr-AdS black hole considered here, the near-horizon analysis above strongly suggests that we should expect pole-skipping whenever $\delta g_{vv}(r_0, x, \psi)$ obeys \eqref{eq:NHgvveq}. However in order to demonstrate that this is indeed the case, it is necessary to perform a more sophisticated of the linearised metric perturbations than we have presented so far. This is because the radial evolution of this near-horizon solution to the boundary is significantly more complicated for Kerr-AdS than for planar or BTZ black holes, due to the difficulty in separating the dependence of the metric perturbations on $r$ and $x$. We will do this in the following Section using the Teukolsky formalism, and show that the intuition suggested by the above argument is correct.

\section{Pole-skipping and quasinormal modes in Kerr-AdS}
\label{sec:QNMSection}

\paragraph{}In the previous Section we saw that when the near-horizon metric had a certain spatial profile related to the shock wave equation, one component of the Einstein equations became trivial at the outer horizon. In the static case, for which the radial and spatial dependence of the metric perturbations can easily be separated, it is simple to then perturbatively construct solutions to the linearised Einstein equations around the horizon and find the extra free parameter in the ingoing solution that is the origin of pole-skipping \cite{Blake:2018leo}. 

\paragraph{} In this Section we will demonstrate that the same is true for Kerr-AdS, and that there is an extra ingoing mode whenever the angular profile of $\delta g_{vv}(r_0, x, \psi)$ satisfies~\eqref{eq:NHgvveq}. As previously mentioned, for Kerr-AdS it is non-trivial to separate variables in the perturbation equations and so to achieve this we will use the Teukolsky formalism: by means of a suitable ansatz for the metric in terms of certain master fields, a separation of variables can be achieved reducing these PDEs to a set of ODEs called the Teukolsky equations \cite{Teukolsky:1972my,Teukolsky:1973ha} (see \cite{Chandrasekhar:1985kt,Dias:2009ex,Dias:2013sdc} for helpful overviews). Using these we will be able to provide an analysis of pole-skipping in terms of ODEs that mirrors previous treatments. In particular this will allow us to explicitly construct the extra ingoing solution at $\omega_*$ whenever the angular profile of $\delta g_{vv}$ at the horizon satisfies~\eqref{eq:NHgvveq}, and to show that its existence implies the usual signatures of pole-skipping such as the existence of a corresponding gravitational quasinormal mode.\footnote{Note that throughout we will use ``quasinormal mode'' in a slightly more general sense than is commonly used -- in particular we will include solutions to the linearised Einstein equations that are asymptotically global AdS$_4$, but whose angular profile in the bulk may not be regular.}

\subsection{The Teukolsky equations}
  
\paragraph{} In this Section we will introduce the key features of the Teukolsky equations that will be needed to address the question of pole-skipping in Kerr-AdS. A detailed discussion of the use of the Teukolsky equations to study quasinormal modes in this spacetime was presented in \cite{Dias:2013sdc}, to which we refer the reader for the many additional details of the results quoted below. 

\paragraph{}For ease of comparison with \cite{Dias:2013sdc}, it will be convenient to use the coordinate system of \cite{Chambers:1994ap} in which the Kerr-AdS line element is
\begin{equation}
\begin{aligned}
ds^2=&\,-\frac{\Delta_r}{\Xi^2\Sigma}\left(d\tilde{t}-\frac{(a^2-\chi^2)}{a}d\phi\right)^2+\frac{\Delta_\chi}{\Xi^2\Sigma}\left(d\tilde{t}-\frac{(a^2+r^2)}{a}d\phi\right)^2+\frac{\Sigma}{\Delta_r}dr^2+\frac{\Sigma}{\Delta_\chi}d\chi^2,
\end{aligned}
\end{equation}
where\footnote{Note that $\Sigma$ here coincides with our previous definition \eqref{eq:BGfunctions} after changing coordinates according to equation \eqref{eq:chidefn}.}
\begin{equation}
\begin{aligned}
\Delta_\chi=(a^2-\chi^2)\left(1-\frac{\chi^2}{L^2}\right),\quad\quad\quad\quad\quad\quad\Sigma=r^2+\chi^2,
\end{aligned}
\end{equation}
and $\Delta_r$ and $\Xi$ are given in equation \eqref{eq:BGfunctions}. The temporal and angular coordinates $\tilde{t}$ and $\chi$ are related to the coordinates used previously by
\begin{equation}
\label{eq:chidefn}
\tilde{t}=\Xi t,\quad\quad\quad\quad\quad \chi=a\cos\theta=ax.
\end{equation}
\paragraph{} The Teukolsky equations allow one to construct solutions to the linearised Einstein equations from radial and angular master field variables \cite{COHEN19755,Chrzanowski:1975wv,Kegeles:1979an,Wald:1978vm,Stewart:1978tm}. To do this, one first parameterises metric perturbations in terms of Hertz potentials $\psi_H^{(\pm2)}$ as
\begin{equation}
\begin{aligned}
\label{eq:Hertzmap}
&\,\delta g^-_{\mu\nu}(r,\chi,\tilde{t},\phi)=\left(l_{(\mu}m_{\nu)}\left(\Delta_1\Delta_2+\Delta_3\Delta_4\right)-l_\mu l_\nu\Delta_2\Delta_5-m_\mu m_\nu\Delta_4\Delta_6\right)\psi_H^{-},\\
&\,\delta g^+_{\mu\nu}(r,\chi,\tilde{t},\phi)=\left(n_{(\nu}\bar{m}_{\mu)}\left(\Delta_7\Delta_8+\Delta_9\Delta_{10}\right)-n_\mu n_\nu\Delta_{10}\Delta_{11}-\bar{m}_\mu \bar{m}_\nu\Delta_8\Delta_{12}\right)\psi_H^{+},
\end{aligned}
\end{equation}
where $l^\mu$, $m^\mu$, $\bar{m}^\mu$, $n^\mu$ and the first order differential operators $\Delta_n$ are defined in Appendix~\ref{app:teukolsky}. Then for Hertz potentials of the form
\begin{equation}
\begin{aligned}
\label{eq:Hertzpotentials}
\psi_H^{\pm}\left(r,\chi, \tilde{t},\phi\right)&\,=e^{-i\tilde{\omega}\tilde{t}+ik\phi}\left(r-i\chi\right)^2R^{\pm}_{\tilde{\omega} k \lambda_{\pm}}(r)S^{\pm}_{\tilde{\omega} k \lambda_{\pm}}(\chi),
\end{aligned}
\end{equation}
one finds that the metrics in \eqref{eq:Hertzmap} satisfy the linearised Einstein equations when the functions $R^{\pm}_{\tilde{\omega} k \lambda_\pm}(r), S^{\pm}_{\tilde{\omega} k \lambda_\pm}(\chi)$ satisfy pairs of ODEs known as the Teukolsky equations. Here $\lambda_{\pm}$ are separation constants that appear from the separation of the $r$ and $\chi$ dependence via the master fields. There are two pairs of Teukolsky equations, one for each of the variables $\psi_H^{\pm}$. For the variable $\psi_{H}^{+}$ we shall refer to these as the spin +2 Teukolsky equations, which have the form 
\begin{equation}
\begin{aligned}
\label{eq:spinPlusTeukolskyeqns}
&\,\left(\mathcal{D}_{-1}\Delta_r\mathcal{D}_1^\dagger+6\left(\frac{r^2}{L^2}+i\tilde{\omega}\Xi r\right)-\lambda_+\right)R^+_{\tilde{\omega} k \lambda_+}(r)=0,\\
&\,\left(\mathcal{L}_{-1}\Delta_\chi\mathcal{L}_1^\dagger+6\left(\frac{\chi^2}{L^2}-\tilde{\omega}\Xi\chi\right)+\lambda_+\right)S^+_{\tilde{\omega} k \lambda_+}(\chi)=0.
\end{aligned}
\end{equation}
The spin $-2$ Teukolsky equations, relevant for the potential $\psi_H^{-}$, are given analogously by the two ordinary differential equations
\begin{equation}
\begin{aligned}
\label{eq:spinMinusTeukolskyeqns}
&\,\left(\mathcal{D}_{-1}^\dagger\Delta_r\mathcal{D}_1+6\left(\frac{r^2}{L^2}-i\tilde{\omega}\Xi r\right)-\lambda_-\right)R^-_{\tilde{\omega} k \lambda_-}(r)=0,\\
&\,\left(\mathcal{L}_{-1}^\dagger\Delta_\chi\mathcal{L}_1+6\left(\frac{\chi^2}{L^2}+\tilde{\omega}\Xi\chi\right)+\lambda_-\right)S^-_{\tilde{\omega} k \lambda_-}(\chi)=0.
\end{aligned}
\end{equation}
The objects $\mathcal{D}_n$, $\mathcal{D}^\dagger_n$, $\mathcal{L}_n$ and $\mathcal{L}^\dagger_n$ are differential operators whose explicit forms are given in Appendix~\ref{app:teukolsky}.
\paragraph{}Solutions to the angular Teukolsky equations characterise the angular dependence of the metric. Whilst solutions for $S^{\pm}_{\tilde{\omega} k \lambda_\pm }(\chi)$ can be obtained for any value of $\lambda_{\pm}$, normally one is interested in angular eigenfunctions for which the resulting metric is regular at both $\chi = \pm a$. This regularity implies $\lambda_+ = \lambda_- =\lambda$ takes a discrete set of values, and the corresponding regular eigenfunctions are called spin-weighted AdS-spheroidal harmonics. For a given $\tilde{\omega}$ and integer $k$ the eigenvalues $\lambda$ at which regular solutions exist can be indexed by an integer $l$, where regularity constrains $-l \leq k \leq l$ and $l \geq 2$. In general, $\lambda(\tilde{\omega}, k, l)$ must be computed numerically, as was done in \cite{Cardoso:2013pza}. However analytic formulae exist in certain limits. In particular, in the slowly rotating limit (i.e.~in an expansion at small $a$) \cite{Cardoso:2013pza,Berti:2005gp,BreuerRA}
\begin{equation}
\lambda(\tilde{\omega}, k, l) = (l-1)(l+2)  - \frac{2 k}{l}\frac{l^2 + l + 4}{l + 1} a \tilde{\omega} + \dots.
\end{equation}

\paragraph{} A crucial property of the Teukolsky equations is that given a solution to one pair of these equations, one can generate a solution to the other pair using differential maps named the Starobinsky-Teukolsky identities \cite{Starobinsky:1973aij,StarobinskyOther,Teukolsky:1974yv,Chand1,Chand2}. If $R^+_{\tilde{\omega} k \lambda}(r)$ and $S^+_{\tilde{\omega} k \lambda}(\chi)$ are solutions to the spin $+2$ Teukolsky equations, then
\begin{equation}
\label{eq:STidentity1}
\mathcal{D}_{-1}^\dagger\Delta_r\mathcal{D}_0^\dagger\mathcal{D}_0^\dagger\Delta_r\mathcal{D}_1^\dagger R^+_{\tilde{\omega} k \lambda}(r)\quad\quad\quad\text{and}\quad\quad\quad\mathcal{L}_{-1}^\dagger\Delta_\chi\mathcal{L}_0^\dagger\mathcal{L}_0^\dagger\Delta_\chi\mathcal{L}_1^\dagger S^+_{\tilde{\omega} k \lambda}(\chi)
\end{equation}
satisfy the spin $-2$ Teukolsky equations for the same values of the constants $\tilde{\omega}$, $k$ and $\lambda_+=\lambda_-=\lambda$. Similarly, if $R^-_{\tilde{\omega} k \lambda}(r)$ and $S^-_{\tilde{\omega} k \lambda}(\chi)$ are solutions to the spin $-2$ Teukolsky equations, then
\begin{equation}
\label{eq:STidentity2}
\mathcal{D}_{-1}\Delta_r\mathcal{D}_0\mathcal{D}_0\Delta_r\mathcal{D}_1 R^-_{\tilde{\omega} k \lambda}(r)\quad\quad\quad\text{and}\quad\quad\quad\mathcal{L}_{-1}\Delta_\chi\mathcal{L}_0\mathcal{L}_0\Delta_\chi\mathcal{L}_1 S^-_{\tilde{\omega} k \lambda}(\chi)
\end{equation}
are solutions to the spin $+2$ equations for the same values of the constants. Successive application of two Starobinsky-Teukolsky identities therefore provides an eighth order differential map between solutions of a single Teukolsky equation. Evaluating such a map explicitly and using the Teukolsky equation to simplify yields
\begin{equation}
\begin{aligned}
\label{eq:eq:STidentity3}
&\,\mathcal{D}_{-1}\Delta_r\mathcal{D}_0\mathcal{D}_0\Delta_r\mathcal{D}_1\mathcal{D}_{-1}^\dagger\Delta_r\mathcal{D}_0^\dagger\mathcal{D}_0^\dagger\Delta_r\mathcal{D}_1^\dagger R^+_{\tilde{\omega} k \lambda}(r)=\mathcal{C}_{ST}^2 R^+_{\tilde{\omega} k \lambda}(r),\\
&\,\mathcal{L}_{-1}\Delta_\chi\mathcal{L}_0\mathcal{L}_0\Delta_\chi\mathcal{L}_1\mathcal{L}_{-1}^\dagger\Delta_\chi\mathcal{L}_0^\dagger\mathcal{L}_0^\dagger\Delta_\chi\mathcal{L}_1^\dagger S^+_{\tilde{\omega} k \lambda}(\chi)=\mathcal{K}_{ST}^2 S^+_{\tilde{\omega} k \lambda}(\chi),
\end{aligned}
\end{equation}
where we have defined Starobinsky-Teukolsky constants
\begin{equation}
\begin{aligned}
\label{eq:STconstantsdefns}
\mathcal{K}_{ST}^2\,&=\lambda^2\left(\lambda+2\right)^2+8\lambda\Xi^2 a\tilde{\omega}\left((6+5\lambda)(k-a\tilde{\omega})+12a\tilde{\omega}\right)+144\Xi^4a^2\tilde{\omega}^2(k-a\tilde{\omega})^2\\
&\,+\frac{4a^2}{L^2}\left[\lambda(\lambda+2)(\lambda-6)+12\Xi^2(k-a\tilde{\omega})(2k\lambda-a\tilde{\omega}(\lambda-6))+\frac{a^2}{L^2}(\lambda-6)^2\right],\\
\mathcal{C}_{ST}^2\,&=\mathcal{K}_{ST}^2+\left(12M\Xi\tilde{\omega}\right)^2.
\end{aligned}
\end{equation}  
  
\subsection{Existence of extra ingoing modes}
\label{sec:TeukolskyQNMs}

\paragraph{} We are now going to take advantage of the Teukolsky equations and associated formalism to prove rigorously our claim from Section \ref{sec:HorizonPoleSkippingAnalysis} that there is an additional ingoing solution for $\omega_* = i 2\pi T$ when the angular profile of the near-horizon metric satisfies equation~\eqref{eq:NHgvveq}. 

\paragraph{}In particular, for our purposes the key point of the Teukolsky equations is they provide an ansatz in which the study of the radial profiles of metric perturbations can be reduced to studying a pair of ordinary differential equations for the radial functions $R^{\pm}_{\tilde{\omega} k \lambda}(r)$ defined above. In particular, for a given value of $(\tilde{\omega}, k, \lambda)$ we will consider metric perturbations of the form 
\begin{equation}
\label{eq:HertzSum}
\delta g_{\mu\nu}(r,\chi,\tilde{t},\phi)=\delta g_{\mu\nu}^{+}(r,\chi,\tilde{t},\phi)+\delta g_{\mu\nu}^{-}(r,\chi,\tilde{t},\phi),
\end{equation}
that are a sum of two solutions generated by the Hertz map \eqref{eq:Hertzmap}. In the case where one imposes regularity in the angular coordinate (i.e.~where $\lambda = \lambda(\tilde{\omega}, k, l)$) then the angular profiles are related to the spin weighted AdS-spheroidal harmonics. For $l \geq 2$, the above ansatz \eqref{eq:HertzSum} is then, up to gauge transformations, the most general solution to the linearised Einstein equations. 

\paragraph{} Here, motivated by previous analyses of pole-skipping, we will consider a more general class of metric perturbations in which we relax the condition of regularity in the angular profile of the metric. To do this, we will consider solutions to the Teukolsky equations where $k$ and $\lambda_+ = \lambda_- = \lambda$ are taken to be arbitrary complex parameters. The metric is then taken to be generated by the Hertz map \eqref{eq:HertzSum} above, where $S^{+}_{\tilde{\omega} k\lambda}(\chi)$ is an arbitrary solution to the spin $+2$ angular Teukolsky equation and $S^{-}_{\tilde{\omega} k\lambda}(\chi)$ is the corresponding solution to the spin $-2$ angular equation obtained by applying the Starobinsky-Teukolsky identity \eqref{eq:STidentity1} to $S^{+}_{\tilde{\omega} k\lambda}(\chi)$. Without loss of generality, we fix the relative normalisations of the two angular solutions by $\mathcal{K}_{ST}S^{-}_{\tilde{\omega} k\lambda}(\chi)=\mathcal{L}_{-1}^\dagger\Delta_\chi\mathcal{L}_0^\dagger\mathcal{L}_0^\dagger\Delta_\chi\mathcal{L}_1^\dagger S^+_{\tilde{\omega} k \lambda}(\chi)$. Whilst considering solutions without spatial regularity may seem unusual, it is analogous to what is done for planar black holes. In that case, the relation between hydrodynamics and chaos becomes manifest upon studying pole-skipping after analytically continuing a real wavevector $\vec{q}$ to a complex one \cite{Grozdanov:2017ajz,Blake:2017ris,Blake:2018leo}.
\paragraph{} Having fixed the angular dependence of the Hertz potentials as above, the metric perturbation \eqref{eq:HertzSum} is characterised by two radial variables $R^{\pm}_{\tilde{\omega} k \lambda}(r)$, each of which satisfies a second order ODE. Here, we are interested in solutions to the Teukolsky equations for which the metric constructed via the Hertz map is ingoing at the outer horizon of the black holes, i.e. for which $\delta g_{\mu\nu}(r,v,\chi,\psi)$ has a Taylor series expansion around the outer horizon (where $v$ is the ingoing coordinate defined in \eqref{eq:nullcoordsdefn}). For generic values of $\tilde{\omega}$, each radial Teukolsky equation has two types of solution near the outer horizon
\begin{equation}
\begin{aligned}
\label{eq:nearhorizonBCsonRgeneric}
&\,R_{\tilde{\omega} k\lambda}^{\pm}(r\rightarrow r_0)\sim (r-r_0)^{\mp1-i\frac{(\Xi\tilde{\omega}-k\Omega_H)}{4\pi T}},\quad\quad\text{and}\\
&\,R_{\tilde{\omega} k\lambda}^{\pm}(r\rightarrow r_0)\sim (r-r_0)^{\pm1+i\frac{(\Xi\tilde{\omega}-k\Omega_H)}{4\pi T}}.
\end{aligned}
\end{equation}
It is the first of these that typically corresponds to an ingoing solution for metric perturbations following the Hertz map, while the second corresponds to an outgoing solution. Therefore after imposing ingoing boundary conditions, we have only one independent solution for each of $R^{\pm}_{\tilde{\omega} k\lambda}$, and so for generic values of $\tilde{\omega}$ ingoing solutions to the Teukolsky equations are parameterised by two constants. 

\paragraph{} Here we will show however that if we treat $(\tilde{\omega}, k, \lambda)$ as arbitrary complex parameters then there are certain values of them for which there exists an extra ingoing mode. To understand how this can be possible, we note that whilst the above discussion holds for a generic frequency, the situations is more subtle is we look at frequencies
\begin{equation}
\label{omegastar}
\tilde{\omega}\equiv\tilde{\omega}_*=\frac{k\Omega_H+i2\pi T}{\Xi}.
\end{equation}
Note that in the co-rotating coordinates used in Section~\ref{sec:OTOCsection} and~\ref{sec:HorizonPoleSkippingAnalysis} this corresponds to a frequency $\omega = i 2\pi T$ with respect to boundary time $t$, which is the frequency at which we expect from Section~\ref{sec:HorizonPoleSkippingAnalysis} an extra ingoing mode to exist. It is also the frequency governing the temporal growth of the OTOC in corotating coordinates. 

\paragraph{} The frequency \eqref{omegastar} is not a generic frequency from the point of view of the near-horizon solutions of the radial Teukolsky equations. For modes of this frequency, the radial profiles of the two independent solutions \eqref{eq:nearhorizonBCsonRgeneric} naively differ by an integer power. Suppose the two independent solutions for $R^+_{\tilde{\omega}_* k\lambda}(r)$ at this frequency did indeed have the form
\begin{equation}
R_{\tilde{\omega}_* k\lambda}^{+}(r\rightarrow r_0)\sim (r-r_0)^{\pm1/2},
\end{equation}
near the outer horizon. This would suggest the existence of an extra ingoing solution, as applying the Hertz map to either of these solutions would produce a metric satisfying ingoing boundary conditions. However, in cases where the naive analysis indicates powers differing by an integer, a more thorough analysis typically shows that one of the solutions contains logarithmic terms that violate the ingoing boundary conditions.

\paragraph{} However, in planar black holes, it has been understood that such logarithms do not appear for certain choices of spatial profile of the perturbation \cite{Blake:2019otz} (which in that case correspond to certain wavenumbers). Here, the analogous result is also true. We find that both of the metric solutions generated via the Hertz map from $R^+_{\tilde{\omega}_* k\lambda}(r)$ are indeed ingoing, provided that the angular profile of the metric components is chosen appropriately. For such angular profiles, there is therefore an extra ingoing mode in the metric solution constructed via the Hertz map. Moreover, we will see that this choice of angular profile is precisely such that the metric on the horizon satisfies equation~\eqref{eq:NHgvveq}.   

\paragraph{} To illustrate how this occurs, we begin by examining the near-horizon properties of the solutions $R^+_{\tilde{\omega}_* k\lambda}(r)$. We first assume there is a solution near $r_0$ that is consistent with ingoing boundary conditions for the metric:
\begin{equation}
R^+_{\tilde{\omega}_* k\lambda}\left(r\rightarrow r_0\right)\rightarrow (r-r_0)^{-1/2}\sum_{n=0}^{\infty}R_n(r-r_0)^n.
\end{equation}
Substituting this into the radial Teukolsky equation \eqref{eq:spinPlusTeukolskyeqns} and solving order-by-order in $(r-r_0)$ for the coefficients yields the leading order condition
\begin{equation}
\label{eq:NHequationsforR}
\left(-\lambda+6\left(ir_0\Xi\tilde{\omega}_*+\frac{r_0^2}{L^2}\right)\right)R_0=0,
\end{equation}
along with expressions for $R_{n>2}$ in terms of $R_0$ and $R_1$. Equation \eqref{eq:NHequationsforR} has the solution $R_0=0$, from which we can then construct a one-parameter ($R_1$) family of ingoing solutions. For generic values of $\lambda$ this is the unique ingoing solution (up to overall normalisation) for $R^+_{\tilde{\omega}_* k\lambda}(r)$. However, for solutions with
\begin{equation}
\label{lambdastar}
\lambda=\lambda_*\equiv 6\left(ir_0\Xi\tilde{\omega}_*+\frac{r_0^2}{L^2}\right),
\end{equation}
the condition \eqref{eq:NHequationsforR} is identically satisfied and therefore both $R_0$ and $R_1$ are independent parameters. Hence in this case there are two independent solutions $R^+_{\tilde{\omega}_* k\lambda_*}(r)$ that each generate an ingoing metric solution via the Hertz map. 

\paragraph{} We can do even better than this near-horizon analysis by observing that the Starobinsky-Teukolsky constant $\mathcal{C}_{ST}^2$ vanishes precisely for $\tilde{\omega}=\tilde{\omega}_*$ and $\lambda=\lambda_*$. As explained in Appendix~\ref{app:poleskipping}, this allows us to construct exact expressions for $R^{\pm}_{\tilde{\omega}_* k\lambda_*}(r)$ throughout the entire spacetime:
\begin{equation}
\label{eq:exactRminussoln}
R^{-}_{\tilde{\omega}_* k\lambda_*}(r)=\frac{(r-r_0)^3}{I(r)\Delta_r(r)}\left(\alpha^-+\beta^-\int^r\frac{\Delta_r(\bar{r})I(\bar{r})^2}{(\bar{r}-r_0)^6}d\bar{r}\right),
\end{equation}
and
\begin{equation}
\label{eq:exactRplussoln}
R^{+}_{\tilde{\omega}_* k\lambda_*}(r)=\frac{G(r)I(r)}{\Delta_r(r)}\left(\alpha^++\beta^+\int^r_{r_0}\frac{\Delta_r(\bar{r})}{I(\bar{r})^2G(\bar{r})^2}d\bar{r}\right),
\end{equation}
in terms of the functions
\begin{equation}
\begin{aligned}
I(r)&\,=\left(1-\frac{r_0}{r}\right)^{1/2}\exp\left(\int^r_{r_0}d\bar{r}\left(i\frac{K_r(\bar{r},\tilde{\omega}_*,k)}{\Delta_r(\bar{r})}-\frac{1}{2(\bar{r}-r_0)}+\frac{1}{2\bar{r}}\right)\right),\\
G(r)&\,=c_3r^3+c_2r^2+r+c_0,
\end{aligned}
\end{equation}
where $c_n$ are fixed constants for which we give explicit expressions in Appendix~\ref{app:poleskipping}. 

\paragraph{} As expected the general solutions \eqref{eq:exactRminussoln} and \eqref{eq:exactRplussoln} depend on four arbitrary constants $\alpha^\pm$ and $\beta^\pm$, and so these generate four independent metric solutions upon applying the Hertz map. It is straightforward to verify using the Hertz map that the three solutions parameterised by $\alpha^{\pm}$ and $\beta^+$ all satisfy ingoing boundary conditions at the outer horizon while the solution parameterised by $\beta^-$ does not. As there are typically only two ingoing solutions to the radial Teukolsky equations at a generic value of $\tilde{\omega}$ -- one each for $R^{\pm}_{\tilde{\omega} k\lambda}(r)$ -- we have shown that for any value of $k$ there is an additional ingoing solution for the case $\tilde{\omega}=\tilde{\omega}_*$ and $\lambda=\lambda_*$.
 
\paragraph{} Finally, we can now ask what the spatial profile of the metric component $\delta g_{vv}$ is at the points $(\tilde{\omega}_*, k , \lambda_*)$ at which the extra ingoing modes exist. To do this we note that after applying the Hertz map the contributions to $\delta g_{vv}$ from the ingoing modes $R^{\pm}_{\tilde{\omega}_* k \lambda_*}$ generated by the constants $\alpha^-, \beta^+$ vanish at the horizon.  In contrast, the mode generated by $\alpha^+$ gives a non-zero contribution at the horizon that  can be expressed in terms of the angular profile $S^+_{\tilde{\omega}_* k \lambda_*}(\chi)$ and its first and second derivatives. Using that $S^+_{\tilde{\omega}_* k \lambda_*}(\chi)$ is a solution to the spin +2 angular Teukolsky equation for $(\tilde{\omega}_*, k,  \lambda_*)$ we then find that parameterising the resulting metric as $\delta g_{vv}(r_0,\chi, \phi) \propto(r_0^2+\chi^2)F(\chi) $ the angular profile $F(\chi)$ satisfies %
\begin{equation}
\begin{aligned}
&\,\partial_\chi\left(\Delta_\chi\partial_\chi F\right)+\frac{a^2K_0^2\Sigma_0^2}{\Delta_\chi}\left(-k^2+2ik\frac{(a^2-\chi^2)}{aK_0\Sigma_0}\left(\alpha+\frac{r_0\Delta_\chi}{(a^2-\chi^2)\Sigma_0}\right)\right)F\\
&\,-\alpha\left(2r_0-\frac{\alpha(a^2-\chi^2)}{1-\frac{\chi^2}{L^2}}\right)F = 0,
\end{aligned}
\end{equation}
where we denote $\Sigma_0=r_0^2+\chi^2$. Note that $\Delta_\chi=(1-x^2)\Delta_x$ and thus after identifying $F=f_1$, $\chi=ax$ and $\partial_\psi= ik$, this is precisely the sourceless version of the shock wave equation \eqref{horizon}.  In summary, we see that the extra ingoing mode in the metric we have found using the Teukolsky equations exists precisely at a frequency $\omega = i 2 \pi T$ when the angular profile of $\delta g_{vv}$ at the horizon is given by a solution of the shock wave equation~\eqref{eq:NHgvveq}. 

\paragraph{} Whilst we had anticipated that this should be the case from Section~\ref{sec:HorizonPoleSkippingAnalysis}, the Teukolsky formalism has allowed us to both demonstrate this rigorously and also provides a characterisation of the spatial profile of the metric everywhere in the bulk through the angular Teukolsky equations. In the next subsection we will discuss the implications of this additional ingoing mode for pole-skipping and the connection between quasinormal modes and the form of the OTOC in the slowly-rotating limit.

\subsection{Quasinormal modes and pole-skipping}

\paragraph{}We have used the Teukolsky formalism to identify an extra ingoing solution to the linearised Einstein equations when the frequency $\tilde{\omega}$ and separation constant $\lambda$ in the Teukolsky formalism have certain values given by equations \eqref{omegastar} and \eqref{lambdastar}. We will now discuss the implications of this extra ingoing mode. In particular we will argue that the existence of the extra ingoing mode means that the spacetime supports a quasinormal mode at the points $(\tilde{\omega}_*, k, \lambda_*)$ for any $k$. More precisely, the boundary energy density response is undefined at these points and supports a collective excitation arbitrarily close to them. This is what is meant by pole-skipping. Note that having used the Teukolsky formalism to relate the radial dependence of gravitational perturbations to the study of two ODEs, our subsequent analysis closely follows previous treatments of pole-skipping \cite{Blake:2018leo} and hence many technical details are relegated to Appendix~\ref{app:poleskipping}. 

\paragraph{} Before proceeding, we reiterate that we are analysing gravitational perturbations for which $(\tilde{\omega}, k, \lambda)$ are taken as independent complex parameters.\footnote{Note for any $(\tilde{\omega}, k, \lambda)$ there are two independent solutions to the angular Teukolsky equation. However the radial solutions $R^{\pm}_{\tilde{\omega} k \lambda}$ depend only on the parameters $(\tilde{\omega}, k, \lambda)$.} In doing so we are not imposing the condition of regularity of the angular profile of the metric, which as discussed above restricts $k$ to an integer and $\lambda$ to a discrete set of values $\lambda(\tilde{\omega}, k, l)$ where $l\geq2$ is an integer and $|k|\leq l$. In particular, we note that the values $\lambda_*$ in equation \eqref{lambdastar} for which the extra ingoing mode exists do not generically correspond to a situation where the bulk metric is regular in angular coordinates. Whilst such solutions to the Einstein equations are therefore not directly physical, as in the planar case their existence implies a connection between hydrodynamics and many-body chaos, and can be used to extract information about the physical dispersion relations of collective excitations in the boundary theory. 

\paragraph{} To explore the implications of the extra ingoing mode, we first discuss the asymptotics of the solutions to the radial Teukolsky equations. For any value of $(\tilde{\omega}, k, \lambda)$, these can be expanded as $r \to \infty$ in the form 
\begin{equation}
R^{\pm}_{\tilde{\omega} k\lambda}(r\rightarrow\infty)\rightarrow A_\pm\frac{L}{r}+B_\pm\frac{L^2}{r^2}+\ldots,
\label{eq:uvexp}
\end{equation}
where $A_\pm, B_\pm$ are constants that depend on $(\tilde{\omega}, k, \lambda)$. Imposing asymptotic boundary conditions on the metric perturbations corresponds to imposing constraints on these constants. In order for the background metric plus the metric perturbation \eqref{eq:HertzSum} to be asymptotically global AdS$_4$, we require that the solutions to the Teukolsky equations satisfy the two conditions \cite{Dias:2013sdc}
\begin{equation}
\begin{aligned}
\label{eq:QNMcond1}
&\,WA_-=LB_-\Bigl[-4akL\Xi\left(2iB_++5L\Xi\tilde{\omega} A_+\right)+2a^2\left(-6+\lambda+6L^2\Xi^2\tilde{\omega}^2\right)A_+\\
&\,+\lambda L^2\left(2+\lambda-4L^2\Xi^2\tilde{\omega}^2\right)A_+-4L^3\Xi\tilde{\omega}\left(iB_++L\tilde{\omega}\Xi A_+\right)\left(2+\lambda-2L^2\Xi^2\tilde{\omega}^2\right)\Bigr],
\end{aligned}
\end{equation}
and
\begin{equation}
\begin{aligned}
\label{eq:QNMcond2}
\mathcal{K}_{ST}=-\frac{B_{-}}{L^3W}&\,\Bigl[\lambda^2L^6(2+\lambda)^2+8a\lambda\left(6+5\lambda\right)kL^6\Xi^2\tilde{\omega}-144a^3kL^4\Xi^2\tilde{\omega}\left(-2+\lambda+2L^2\Xi^2\tilde{\omega}^2\right)\\
&\,+4a^2L^4\left\{\lambda\left(-12+(-4+\lambda)\lambda+24k^2\right)+2L^2\Xi^2\tilde{\omega}^2\left((6-5\lambda)\lambda+18k^2\right)\right\}\\
&\,+4a^4L^2\left\{36-12\lambda+\lambda^2-48\lambda k^2+12L^2\Xi^2\tilde{\omega}^2\left(\lambda-6(1+k^2)\right)+36L^4\Xi^4\tilde{\omega}^2\right\}\\
&\,+48a^6k^2\left(2\lambda+3L^2\Xi^2\tilde{\omega}^2\right)\Bigr],
\end{aligned}
\end{equation}
where the constant $W$ is
\begin{equation}
\begin{aligned}
W=&\,L^3\left[\lambda\left(2+\lambda\right)B_+-4\left(1+\lambda\right)L\tilde{\omega}\Xi\left(i\lambda A_++2L\Xi\tilde{\omega}B_+\right)+4i\left(2+3\lambda\right)L^3\Xi^3\tilde{\omega}^3A_+\right]\\
&\,+8L^7\Xi^4\tilde{\omega}^4\left(B_+-iL\Xi\tilde{\omega}A_+\right)-4akL^2\Xi\left[3i\lambda A_++L\Xi\tilde{\omega}\left(5B_+-8iL\Xi\tilde{\omega}A_+\right)\right]\\
&\,+2a^2L\left[2\left(2iL\Xi\tilde{\omega}A_+-B_+\right)\left(3+\lambda-3L^2\Xi^2\tilde{\omega}^2\right)+3B_+\right].
\end{aligned}
\end{equation}
If, for a given value of $(\tilde{\omega}, k, \lambda)$, the ingoing solution is such that \eqref{eq:QNMcond1} and \eqref{eq:QNMcond2} are satisfied, then the bulk spacetime supports a quasinormal mode. As noted previously, we are using the term quasinormal mode in a more general sense than normal, since we are not demanding regularity of the angular profile of the metric. 
\paragraph{}For generic values of $(\tilde{\omega}, k, \lambda)$, it is simple to see that there will not exist a quasinormal mode. In particular, as discussed in Section~\ref{sec:TeukolskyQNMs}, generically there are unique (up to normalization) ingoing solutions for each of $R^\pm_{\tilde{\omega} k \lambda}(r)$. These ingoing solutions determine the ratios $B_+/A_+$ and $ B_-/A_-$ in terms of $(\tilde{\omega}, k, \lambda)$. The resulting conditions \eqref{eq:QNMcond1} and \eqref{eq:QNMcond2} then become two equations for two free parameters (which can be taken to be $A_{\pm}$). Hence for generic $(\tilde{\omega}, k, \lambda)$ there will not be a non-trivial solution to \eqref{eq:QNMcond1} and  \eqref{eq:QNMcond2} for $A_{\pm}$ and hence there will not be a quasinormal mode. Rather, quasinormal modes only exist for certain frequencies $\tilde{\omega}(k, \lambda)$. 
\paragraph{} However, it is immediately clear that the existence of the extra ingoing mode means that there must be a quasinormal mode for the special case $(\tilde{\omega}_*,k, \lambda_*)$. In this case only the ratio $B_-/A_-$ is fixed by the ingoing boundary condition, and $(A_+, B_+)$ are independent parameters. As such there are three free parameters in the expansion~\eqref{eq:uvexp}, which one expects can be chosen to satisfy the two constraints in \eqref{eq:QNMcond1} and \eqref{eq:QNMcond2}.  Indeed, we can explicitly check this can be achieved by using the ingoing solutions identified in equations~\eqref{eq:exactRminussoln} and \eqref{eq:exactRplussoln}. After first imposing ingoing boundary conditions by setting $\beta^-=0$ in equations \eqref{eq:exactRminussoln} and \eqref{eq:exactRplussoln}, we can expand the resulting solutions near the asymptotic boundary. We then find that the conditions \eqref{eq:QNMcond1} and \eqref{eq:QNMcond2} can be explicitly satisfied for certain choices of $\beta_+/\alpha_+$ and $\beta_+/\alpha_-$, which we provide in Appendix~\ref{app:poleskipping}. By choosing these values of the constants and then applying the Hertz map to the solution in \eqref{eq:exactRminussoln} and \eqref{eq:exactRplussoln}, we therefore explicitly obtain a quasinormal mode solution to the linearised Einstein equations.

\paragraph{}Whilst we have seen there is a quasinormal mode at the special point $(\tilde{\omega}_*,k, \lambda_*)$, there are also other ingoing solutions we could construct at this point, corresponding to any other choice of the ratios $\beta_+/\alpha_+$ and $\beta_+/\alpha_-$. Such solutions will not give rise to metrics that are asymptotically global AdS$_4$, but rather solutions in which the metric of the boundary field theory is non-zero (i.e.~solutions for which there is a source for the stress tensor). The existence of these multiple ingoing solutions means that strictly at the location $(\tilde{\omega}_*,k, \lambda_*)$, the energy density response for a given source in the boundary theory is undefined, i.e.~there is pole-skipping.

\paragraph{}  In order to make the source and energy density response well defined, it is necessary to move slightly away from the pole-skipping point. In particular, continuing to treat $\lambda$ as a free parameter, we can now examine what happens when one moves slightly away from the special point $(\tilde{\omega}_*,k,\lambda_*)$ to
\begin{equation}
\label{eq:pertaway}
\tilde{\omega}=\tilde{\omega}_*+\epsilon\delta\tilde{\omega},\quad\quad\quad\quad \lambda=\lambda_*+\epsilon\delta\lambda,
\end{equation}
where $\epsilon\ll 1$. Since we are no longer exactly at the pole-skipping point, there is no longer an extra ingoing solution. This can be seen explicitly by including non-zero $\epsilon$ corrections to the horizon equation~\eqref{eq:NHequationsforR} from which one constructs the ingoing solutions to $R^+_{\tilde{\omega} k \lambda}(r)$. At order $\epsilon$ this equation receives corrections proportional to $\delta \tilde{\omega}, \delta \lambda$ that determine the ratio $R_1/R_0$ in terms of $\delta \tilde{\omega}/\delta \lambda$. Whilst the ingoing solution now only has two free parameters, the fact the solutions depend on the slope $\delta \tilde{\omega}/\delta \lambda$ means that in the vicinity of the pole-skipping point we can generate a one-parameter family of ingoing solutions of the form \eqref{eq:exactRminussoln} and \eqref{eq:exactRplussoln} by varying the slope. In particular, as shown in Appendix~\ref{app:poleskipping}, the slope $\delta \tilde{\omega}/\delta \lambda$ can be chosen such that the resulting ingoing solution near $(\tilde{\omega}_*, k, \lambda_*)$ corresponds to a quasinormal mode. Thus, as for planar black holes, there must be a quasinormal mode whose dispersion relation $\tilde{\omega}(k,\lambda)$ passes through the pole-skipping point $(\tilde{\omega}_*, k, \lambda_*)$.

\subsection{Equatorial OTOC and pole-skipping in the slowly rotating limit} 
\label{sec:QNMconstraints}

\paragraph{} In the previous subsections we showed that there must exist a quasinormal mode in the Kerr-AdS spacetime when the parameters in the Teukolsky equations are taken to be $(\tilde{\omega}_*, k, \lambda_*)$. We also observed a connection between the extra ingoing mode and many-body quantum chaos, specifically that such extra ingoing modes exist when the angular profile of the metric component $\delta g_{vv}$ on the horizon satisfies the shock wave equation~\ref{eq:NHgvveq}. 

\paragraph{} The above connection is non-trivial and, as in planar and non-rotating cases, provides strong evidence that a hydrodynamic effective theory of chaos continues to apply in Kerr-AdS black holes. However it would also be nice to obtain a more direct connection between hydrodynamic and chaotic observables in the boundary quantum field theory. In particular, in planar black holes, there is a simple and general relationship between the functional form of the OTOC and the locations of pole-skipping points \cite{Grozdanov:2017ajz,Blake:2017ris,Blake:2018leo}. Here we will explain how such a connection generalises to the Kerr-AdS black hole, at least for large black holes in the slowly rotating limit for which we computed the equatorial OTOC in Section~\ref{sec:OTOCsection}. 

\paragraph{} In particular, let us recall that in the Kerr-AdS black hole, we have found a family of pole-skipping points when the parameters $(\tilde{\omega}, k, \lambda)$ in the Teukolsky equations satisfy
\begin{equation}
\tilde{\omega}_* = \frac{k \Omega_H + i 2 \pi T}{\Xi},
\end{equation}
and 
\begin{equation}
\lambda_* = 6 \bigg( i r_0 \Xi \tilde{\omega_*} + \frac{r_0^2}{L^2} \bigg),
\end{equation} 
for any value of $k$.

\paragraph{} We wish to compare the locations of these pole-skipping points to the profile of the equatorial OTOC~\eqref{equatorialOTOC} for large black holes in the slowly rotating limit. For comparison to the pole-skipping analysis, it is instructive to write this functional form in co-rotating coordinates by defining $\psi = \Phi - \Omega t$. We then have 
\begin{equation}
H_4(t, \psi) \propto\frac{e^{2 \pi T t}}{\sqrt{|\nu_0\psi|}}\begin{cases} \exp\left(-\left(1+\frac{\sqrt{3}a}{2L}\right) \nu_0\psi \right), & \psi>0,\\ \exp\left(\left(1-\frac{\sqrt{3}a}{2L}\right) \nu_0\psi \right), & \psi<0.\end{cases}
\label{otoccomp}
\end{equation}
In co-rotating coordinates, the equatorial OTOC is therefore characterised by exponential growth with a frequency $\omega_* = i 2\pi T$, and two imaginary wavenumbers $k = \pm i k_\pm$ where
\begin{equation}
k_\pm = \nu_0 \left(1 \pm \frac{\sqrt{3} a}{2 L}  \right) = \frac{2 \pi T_\pm}{v_B^\pm} .
\label{otocwavevector}
\end{equation}
\paragraph{} As we noted below equation~\eqref{omegastar}, the value of $\tilde{\omega}_*$ at which there exists pole-skipping corresponds in co-rotating coordinates to the frequency $\omega_*= i 2\pi T$. As such pole-skipping occurs precisely at the same imaginary frequency that characterises the time dependence of the OTOC~\eqref{otoccomp}. However it is not immediately clear how the wavenumbers $k = \pm i k_\pm$ that characterise the angular dependence of the equatorial OTOC are related to the pole-skipping points, which exist for any value of $k$ provided $\lambda$ is suitably chosen. 

\paragraph{} In order to proceed, it is necessary to understand how $\lambda$ and $k$ should be related when comparing pole-skipping to the equatorial OTOC.  In order to do this we recall that if one is interested in regular solutions to the Einstein equations, then $\lambda$ is not an arbitrary free parameter. Rather, regular solutions exist for values $\lambda(\tilde{\omega}, k, l)$ where $k, l$ are integers with $l\geq2$ and $|k| \leq l$. For a slowly rotating black hole the values of $\lambda$ are given by  \cite{Cardoso:2013pza,Berti:2005gp,BreuerRA}

\begin{equation}
\lambda(\tilde{\omega}, k, l) = (l-1)(l + 2) - \frac{2 k}{l} \frac{l^2 + l + 4}{l + 1} a \tilde{\omega} + \dots, 
\label{lambdaslow}
\end{equation}
and physically relevant quasinormal modes (i.e. those with regular angular profiles) correspond to frequencies $\omega(k,l)$ at which ingoing solutions to the radial Teukolsky equations with $(\tilde{\omega}, \lambda(\tilde{\omega}, k,l), k)$ satisfy the asymptotic boundary conditions in equations~\eqref{eq:QNMcond1} and \eqref{eq:QNMcond2} (see \cite{Cardoso:2013pza} for a study of these modes and \cite{Garbiso:2020puw} for a higher-dimensional generalisation). Now, for planar black holes, if we wish to relate the functional form of the OTOC along the $x$ direction to the location of pole-skipping points, we would do this by considering pole-skipping points where the wavevector $\vec{q}$ in the boundary Green's function is aligned along the $x$ direction \cite{Grozdanov:2017ajz,Blake:2017ris,Blake:2018leo}. At least in the slowly rotating and large black hole limits, a natural analogue for comparing to the equatorial OTOC in the Kerr-AdS black hole is to consider modes where the wavenumber $k$ takes its maximal value, i.e. we consider modes with $k=l$.\footnote{The results below for the values of $k_\pm$ are unchanged if we instead consider $k=-l$.} The corresponding quasinormal modes are then characterised by dispersion relations $\omega(k)$.  

\paragraph{} As we have emphasised, solutions to the Teukolsky equations at the pole-skipping points $(\tilde{\omega}_*, k, \lambda_*)$ do not typically give rise to bulk metric perturbations with regular angular profiles. However motivated by the above discussion, we can consider pole-skipping points in which $\lambda$, $k$ are complex parameters related by \eqref{lambdaslow} where we set $l =k$.  In the large black hole and slowly rotating limit, such pole-skipping points exist precisely when the wavenumber $k$ takes the values \eqref{otocwavevector} related to the form of the OTOC. To show this is the case, we first set $l =k$ and $\tilde{\omega} = \tilde{\omega}_*$ in~\eqref{lambdaslow}. We then find that to leading order in the large black hole  $r_0/L \gg 1$ and slowly rotating limits $a/L \ll 1$ the condition~\eqref{lambdastar} for $\lambda = \lambda_*$ becomes\footnote{When taking the large black hole limit we take $k \sim r_0/L \gg 1$.}
\begin{equation} 
k^2 - \frac{3 i k a r_0}{L^2} = - \frac{3 r_0^2}{L^2}.
\end{equation}
Solving this equation and expanding in the slowly rotating limit we thus find pole-skipping at the wavenumbers
\begin{equation}
k = \pm i \nu_0 \bigg( 1 \pm \frac{\sqrt{3} a}{2 L} \bigg) = \pm i k_\pm, 
\end{equation}
which precisely match the wavenumbers governing the angular decay of the equatorial OTOC. As such the dispersion relation of collective modes of energy density with $l = k $ in the boundary theory should, if analytically continued as described above, pass through the location $(\tilde{\omega}_*, \pm i k_\pm)$ which is precisely related to the functional form of the equatorial OTOC as in \eqref{otoccomp}. As for the form of the OTOC in this limit, the expressions for the pole-skipping locations can be obtained by applying a boost with velocity $v=-\Omega L\ll1$ to those of the static Schwarzschild-AdS$_4$ black hole found in \cite{Blake:2018leo}. 

\section{Discussion}

\paragraph{} In this paper we have explored the connection between many-body quantum chaos and energy dynamics in the holographic theory dual to the Kerr-AdS black hole. We showed that for perturbations at a frequency $\omega_* = i 2 \pi T$ the energy density response of the boundary theory exhibits the phenomenon of pole-skipping whenever the angular profile of the near-horizon metric obeys the gravitational shock wave equation, which also governs the form of the OTOC. As a consequence we found that the bulk Einstein's equations admit quasi-normal mode solutions with such profiles, corresponding to collective excitations in the energy density response of the boundary quantum field theory. For large black holes in the slowly rotating limit we were able to explicitly determine the form of the equatorial OTOC of the boundary theory, and we showed how the pole-skipping phenomenon relates this functional form to the dispersion relation of collective excitations in the energy density response.

\paragraph{} There are many interesting questions for future work. Most immediately, we saw in Section~\ref{sec:OTOCsection} that determining the full shock wave profile and understanding the general form of the boundary OTOC in Kerr-AdS is significantly more involved than in previous examples. It would be interesting to understand the explicit form of the OTOC beyond the slowly-rotating limit we discussed in Section~\ref{sec:slowrototoc}, and also to understand its behaviour outside the equatorial plane. It would also be interesting to explore if the profile of the OTOC beyond these limits can be used to obtain additional constraints on the dispersion relations of collective excitations in the boundary theory.

\paragraph{} Nevertheless, through our analysis of gravitational shock waves we have established a precise connection between the OTOC and pole-skipping in the energy response of the boundary theory dual to the Kerr-AdS black hole. As such, our results provide strong evidence that many-body quantum chaos in such systems can be described in terms of a hydrodynamic effective theory along the lines of \cite{Blake:2017ris}. It would be worthwhile to generalise such an effective theory to include the effects of a chemical potential for rotation. In particular, it is known that in the absence of such a chemical potential the effective theory discussed in \cite{Blake:2017ris, Blake:2021wqj} is only consistent for maximally chaotic systems. For the rotating black holes considered here and in \cite{Jahnke:2019gxr} the exponential growth of the OTOC cannot simply be characterised through a Lyapunov exponent $\lambda_{L} = 2 \pi T$ saturating the bound of \cite{Maldacena:2015waa}. However, both the equatorial OTOC for slowly rotating Kerr-AdS black holes computed in this paper and the OTOC of the BTZ theory computed in \cite{Jahnke:2019gxr} saturate a generalised version of the chaos bound for rotating ensembles introduced in \cite{Mezei:2019dfv}. It would be interesting to understand if saturating such a bound is a necessary condition for a hydrodynamic effective theory of chaos to apply to systems with a chemical potential for rotation. 

\paragraph{}  Furthermore, in this paper we have only considered one family of pole-skipping points, present at a frequency $\omega_* = i 2 \pi T$ in co-rotating coordinates. In planar black holes, there are in addition infinitely many pole-skipping points in the energy density response at complex frequencies in the lower half plane~\cite{Blake:2019otz}. Whilst these additional pole-skipping points are not directly related to chaos, they can be used to obtain a set of exact constraints on the dispersion relations of quasinormal modes in an analogous manner to our discussion in Section~\ref{sec:QNMconstraints}. In planar cases, some of these pole-skipping points occur at real wave-numbers, and thus by generalising them to Kerr-AdS (and other rotating spacetimes) it may be possible to obtain direct constraints on the dispersion relations of the physical quasi-normal modes that are regular in the angular coordinates. In particular, we note that such an analysis is not limited to asymptotically AdS spacetimes, but could also be carried out for spacetimes that are asymptotically flat. As such the work in this paper provides a starting point to identifying the resulting constraints on the quasi-normal mode spectrum of rotating black holes, including those of astrophysical relevance. Finally, mathematical connections between supersymmetric field theories and the quasi-normal modes of black holes have recently been observed \cite{Aminov:2020yma,Hatsuda:2020iql,Bianchi:2021xpr,Bonelli:2021uvf,Bianchi:2021mft} and it would be interesting to understand the significance of the special class of quasi-normal modes that exist at pole-skipping points in this regard.

\acknowledgments

We are grateful to \'{O}scar Dias and Jorge Santos for helpful correspondence. The work of R.~D.~is supported by the STFC Ernest Rutherford Grant ST/R004455/1.
 
 \newpage
\appendix
\section{Kerr-AdS metric in Kruskal-like coordinates}
\label{app:kruskalcoords}

\paragraph{}In the main text we argued that after a shift in the coordinate $U \to U + f_1(x,\psi)$ across the $V=0$ horizon of the metric~\eqref{kruskal}, then at leading order in $f_1$ the only non-trivial perturbation of the metric is the $\delta g_{VV}$ component given by~\eqref{ansatz}. To justify this, let us note that the AdS-Kerr metric~\eqref{kruskal} in Kruskal coordinates is of the form  
\begin{eqnarray}
ds^2 &=& g_{\mu \nu}(U,V,x,\psi) dx^{\mu} dx^{\nu},
\label{metricapp}
\end{eqnarray}
where $x^{\mu} = (U,V,x,\psi)$ and the non-zero metric components are
\begin{equation}
\begin{aligned}
\label{eq:Kruskalmetricomps}
g_{UU}=&\frac{1}{U^2}\frac{1}{4\alpha^2(r_0^2+a^2)^2\Sigma}\left[\Delta_r\left(\frac{\Sigma^2(r_0^2+a^2)^2}{(r^2+a^2)^2}-(r_0^2+a^2 x^2)^2\right)+a^2(1-x^2)\Delta_x(r_0^2-r^2)^2\right],\\
g_{VV}=&\frac{1}{V^2}\frac{1}{4\alpha^2(r_0^2+a^2)^2\Sigma}\left[\Delta_r\left(\frac{\Sigma^2(r_0^2+a^2)^2}{(r^2+a^2)^2}-(r_0^2+a^2x^2)^2\right)+a^2 (1-x^2)\Delta_x(r_0^2-r^2)^2\right],\\
g_{UV}=&\frac{1}{UV}\frac{1}{4\alpha^2(r_0^2+a^2)^2\Sigma}\left[\Delta_r\left(\frac{\Sigma^2(r_0^2+a^2)^2}{(r^2+a^2)^2}+(r_0^2+a^2x^2)^2\right)-a^2(1-x^2)\Delta_x(r_0^2-r^2)^2\right],\\
g_{U\psi}=&\frac{1}{U}\frac{a(1-x^2)}{2\alpha(r_0^2+a^2)\Xi\Sigma}\left[\Delta_x(r^2+a^2)(r_0^2-r^2)-\Delta_r(r_0^2+a^2x^2)\right],\\
g_{V\psi}=&-\frac{1}{V}\frac{a(1-x^2)}{2\alpha(r_0^2+a^2)\Xi\Sigma}\left[\Delta_x(r^2+a^2)(r_0^2-r^2)-\Delta_r(r_0^2+a^2x^2)\right],\\
g_{\psi\psi}=&\frac{(1-x^2)}{\Xi^2\Sigma}\left[\Delta_x(r^2+a^2)^2-a^2(1-x^2)\Delta_r\right],\\
g_{xx}=&\frac{\Sigma}{(1-x^2)\Delta_x},
\end{aligned}
\end{equation}
and $r = r(U V)$. 
\paragraph{}To construct the shock wave ansatz we follow \cite{Roberts:2014isa, Shenker:2014cwa, Shenker:2013pqa,Sfetsos:1994xa} and let $\tilde{U} = U +\theta(V) f_1(x,\psi)$ and make the replacements
\begin{eqnarray}
g_{\mu \nu}(U, V, x, \psi) &\to& g_{\mu \nu}(\tilde{U}, V, x, \psi),  \nonumber \\
 dU &\to& d\tilde{U}  - f_1(x,\psi) \delta(V)dV,
\end{eqnarray}
in \eqref{metricapp}. In terms of coordinates $y^{\mu} = (\tilde{U}, V, x, \psi)$ the resulting metric is given by 
\begin{eqnarray}
ds^2 &=& g_{\mu \nu}(\tilde{U},V,x,\psi) dy^{\mu} dy^{\nu} + \delta ds^2,
\label{metricapp2}
\end{eqnarray}
where at linear order in $f_1(x,\psi)$ one obtains
\begin{eqnarray}
\delta ds^2 &=& - 2 g_{UV}(\tilde{U}, V, x, \psi) f_1(x, \psi) \delta(V) dV^2 -  2 g_{UU}(\tilde{U}, V, x, \psi) f_1(x, \psi) \delta(V) d\tilde{U} dV \nonumber \\
  &-& 2 g_{U \psi}(\tilde{U}, V, x, \psi) f_1(x, \psi) \delta(V) dV d\psi .
 \end{eqnarray}
Note that near the $V=0$ horizon one has 
\begin{equation}
g_{U V} = \frac{\Sigma_0(x) \Delta_r'(0)}{2 \alpha^2 (r_0^2 + a^2)^2} + {\cal O}(V),  \;\;\;\;\;\;\; g_{U \psi} \sim {\cal O}(V), \;\;\;\;\;\;\; g_{U U} \sim {\cal O}(V^2), \;\;\;\;\;
\end{equation}
where $\Sigma_0(x) = r_0^2 + a^2 x^2$. We therefore find that the shock wave metric is of the form \eqref{ansatz} presented in the main text.

\section{Shock wave profile in slowly-rotating limit}
\label{app:slowlyrotating}

\paragraph{}In this Appendix we provide more details on how to obtain the equatorial shock wave profile \eqref{otoccorotating} in the slowly-rotating, large black hole limit. Starting from the slowly-rotating limit of the shock wave equation \eqref{shockslowly} in real space, the first step is to make a Fourier transform of the homogeneous equation with respect to $\psi$ to obtain
\begin{equation}
\frac{d}{dx}\left[(1-x^2)\frac{d}{dx}f_{0,k}(x,0)\right]-\left(\frac{k^2}{1-x^2}+\left(1+\frac{3r_0^2}{L^2}\right) -3ik\frac{a}{r_0}\left(1+\frac{r_0^2}{L^2}\right)\right)f_{0,k}(x,0)=0.
\end{equation}
The solution of this equation that is regular at the North pole $(x=1)$ is the associated Legendre polynomial $f_{0,k}(x>0,0)=c_+P^k_{-1/2+i\nu}(x)$ where $\nu$ is given in equation \eqref{eq:nudefn} in the main text and $c_+$ is an arbitrary constant. We can then obtain the solution that is regular at the South pole $f_{0,k}(x<0,0)=c_-P^k_{-1/2+i\nu}(-x)$ by using the reflection symmetry of the homogeneous equation. We then find the solution to the equation with a delta function source in the standard way by patching the two solutions together across the equator, yielding (up to an overall normalisation)
\begin{equation}
f_0(0,0,\psi)=\sum_{k\in\mathbb{Z}}e^{ik\psi}\lim_{x\rightarrow0}\frac{P^k_{-1/2+i\nu}(x)}{\frac{d}{dx}P^k_{-1/2+i\nu}(x)}.
\end{equation}
After explicitly evaluating the limit of the ratio of associated Legendre polynomials \cite{AbramowitzStegun}, and using gamma function identities, this simplifies to equation \eqref{otocrotating}.

\paragraph{}We will now invert this Fourier series to obtain a real-space form for the equatorial OTOC. To do so, we take the large black hole limit $r_0\gg L$ which simplifies the arguments of the gamma functions such that
\begin{equation}
f_0(0,0,\psi)=\sum_{k\in\mathbb{Z}}\frac{\Gamma\left[\frac{1}{4}-\frac{k}{2}\left(1+\frac{\sqrt{3}a}{2L}\right)-\frac{i\nu_0}{2}\right]\Gamma\left[\frac{1}{4}-\frac{k}{2}\left(1-\frac{\sqrt{3}a}{2L}\right)+\frac{i\nu_0}{2}\right]}{\Gamma\left[\frac{3}{4}-\frac{k}{2}\left(1+\frac{\sqrt{3}a}{2L}\right)-\frac{i\nu_0}{2}\right]\Gamma\left[\frac{3}{4}-\frac{k}{2}\left(1-\frac{\sqrt{3}a}{2L}\right)+\frac{i\nu_0}{2}\right]}e^{ik\psi},
\end{equation}
where $\nu_0=\sqrt{3}r_0/L$. In the limit $\psi\ll1$ with $\nu_0\psi$ fixed, the Fourier sum can be approximated by the Fourier integral
\begin{equation}
\begin{aligned}
f_0(0,0,\psi)=\int_{-\infty}^{\infty} d\tilde{k}\frac{\Gamma\left[\frac{1}{4}-\frac{\tilde{k}\nu_0}{2}\left(1+\frac{\sqrt{3}a}{2L}\right)-\frac{i\nu_0}{2}\right]\Gamma\left[\frac{1}{4}-\frac{\tilde{k}\nu_0}{2}\left(1-\frac{\sqrt{3}a}{2L}\right)+\frac{i\nu_0}{2}\right]}{\Gamma\left[\frac{3}{4}-\frac{\tilde{k}\nu_0}{2}\left(1+\frac{\sqrt{3}a}{2L}\right)-\frac{i\nu_0}{2}\right]\Gamma\left[\frac{3}{4}-\frac{\tilde{k}\nu_0}{2}\left(1-\frac{\sqrt{3}a}{2L}\right)+\frac{i\nu_0}{2}\right]}e^{i\tilde{k}(\nu_0\psi)}.
\end{aligned}
\end{equation}
In the large black hole limit $\nu_0\gg1$, the Fourier coefficients can be simplified using the large $z$ asymptotic expansion $\Gamma(z+c)/\Gamma(z+b)=z^{c-b}+\ldots$ \cite{AbramowitzStegun} to yield (up to an overall normalisation)
\begin{equation}
\label{eq:InverseFT}
f_0(0,0,\psi)=\int_{-\infty}^{\infty} d\tilde{k}\frac{e^{i\tilde{k}(\nu_0\psi)}}{\sqrt{(\tilde{k}-\tilde{k}_+)(\tilde{k}-\tilde{k}_-)}},
\end{equation}
where the branch points are located on the imaginary axis at $\tilde{k}_\pm=\pm i(1\pm \sqrt{3}a/(2L))$.

\paragraph{}The final step is to evaluate the inverse Fourier transform \eqref{eq:InverseFT}. We first choose the branch cuts to extend out from the branch points in opposite directions along the imaginary axis to infinity. To perform the inverse Fourier transform for $\psi>0$ $(\psi<0)$, we then evaluate the contour integral running along the real axis and extending into the upper (lower) half plane and closing by wrapping around the branch cut. The result of this is 
\begin{equation}
f_0(0,0,\psi)= e^{-\frac{\sqrt{3}a}{2L}\nu_0\psi}\int^\infty_0 dq\frac{e^{-(q+1)|\nu_0\psi|}}{\sqrt{q(q+2)}}=e^{-\frac{\sqrt{3}a}{2L}\nu_0\psi}\int^\infty_0dpe^{-|\nu_0\psi|\cosh(p)},
\end{equation}
where in the second expression we made the substitution $q=\cosh(p)-1$. The integral over $p$ is an integral representation of a modified Bessel function of the second kind $K_0(|\nu_0\psi|)$ \cite{AbramowitzStegun} and thus the shock wave profile in the slowly rotating, large black hole limit can be expressed as
\begin{equation}
f_0(0,0,\psi)\propto e^{-\frac{\sqrt{3}a}{2L}\nu_0\psi}K_0\left(|\nu_0\psi|\right).
\end{equation}
When $a=0$, this agrees with the corresponding solution for the planar Schwarzschild-AdS$_4$ black hole \cite{Roberts:2014isa}. The final step is then to expand the Bessel function far from the delta function source $\left|\nu_0\psi\right|\gg 1$, which yields the expression \eqref{otoccorotating} for the shock wave profile presented in the main text.

\section{Additional details of Teukolsky formalism} 
\label{app:teukolsky}

\paragraph{}For the purposes of completeness, in this Appendix we present explicit expressions for the differential operators that play an important role in the Teukolsky formalism of perturbations of the Kerr-AdS black hole \cite{Dias:2013sdc}. In terms of the function $\Sigma=r^2+\chi^2$, the angular differential operators appearing in the Hertz map \eqref{eq:Hertzmap} are
\begin{equation}
\begin{aligned}
&l^\mu\partial_\mu=\frac{1}{\sqrt{2\Sigma}}\left(\frac{\Xi(a^2+r^2)}{\sqrt{\Delta_r}}\partial_{\tilde{t}}+\sqrt{\Delta_r}\partial_r+\frac{a\Xi}{\sqrt{\Delta_r}}\partial_\phi\right),\\
&n^\mu\partial_\mu=\frac{1}{\sqrt{2\Sigma}}\left(\frac{\Xi(a^2+r^2)}{\sqrt{\Delta_r}}\partial_{\tilde{t}}-\sqrt{\Delta_r}\partial_r+\frac{a\Xi}{\sqrt{\Delta_r}}\partial_\phi\right),\\
&m^\mu\partial_\mu=\frac{1}{\sqrt{2\Sigma}}\left(-\frac{i\Xi(a^2-\chi^2)}{\sqrt{\Delta_\chi}}\partial_{\tilde{t}}+\sqrt{\Delta_\chi}\partial_{\chi}-\frac{ia\Xi}{\sqrt{\Delta_\chi}}\partial_\phi\right),\\
&\bar{m}^\mu\partial_\mu=\frac{1}{\sqrt{2\Sigma}}\left(\frac{i\Xi(a^2-\chi^2)}{\sqrt{\Delta_\chi}}\partial_{\tilde{t}}+\sqrt{\Delta_\chi}\partial_{\chi}+\frac{ia\Xi}{\sqrt{\Delta_\chi}}\partial_\phi\right),
\end{aligned}
\end{equation}
and the differential operators $\Delta_n$ are
\begin{equation}
\begin{aligned}
&\Delta_1=l^\mu\partial_\mu+\frac{1}{\sqrt{2\Sigma\Delta_r}}\left(\Delta_r'-\frac{\Delta_r(2r-i\chi)}{\Sigma}\right), \quad\quad \Delta_2=m^\mu\partial_\mu+\frac{1}{\sqrt{2\Sigma\Delta_\chi}}\left(\dot{\Delta}_\chi+\frac{5i\Delta_\chi}{r-i\chi}\right),\\
&\Delta_3=m^\mu\partial_\mu+\frac{1}{\sqrt{2\Sigma\Delta_\chi}}\left(\dot{\Delta}_\chi-\frac{\Delta_\chi(2\chi+ir)}{\Sigma}\right),\quad \Delta_4=l^\mu\partial_\mu+\frac{1}{\sqrt{2\Sigma\Delta_r}}\left(\Delta_r'-\frac{5\Delta_r}{r-i\chi}\right),\\
&\Delta_5=m^\mu\partial_\mu+\frac{1}{\sqrt{2\Sigma\Delta_\chi}}\left(\frac{\dot{\Delta}_\chi}{2}+\frac{ir\Delta_\chi}{\Sigma}\right),\quad\quad\quad\;\;\;\Delta_6=l^\mu\partial_\mu+\frac{1}{\sqrt{2\Sigma\Delta_r}}\left(\frac{\Delta_r'}{2}-\frac{i\chi\Delta_r}{\Sigma}\right),\\
&\Delta_7=\bar{m}^\mu\partial_\mu+\frac{1}{\sqrt{2\Sigma\Delta_\chi}}\left(\dot{\Delta}_\chi-\frac{\Delta_\chi(2\chi+ir)}{\Sigma}\right),\quad \Delta_8=n^\mu\partial_\mu+\frac{1}{\sqrt{2\Sigma\Delta_r}}\left(-\Delta_r'+\frac{5\Delta_r}{r-i\chi}\right),\\
&\Delta_9=n^\mu\partial_\mu+\frac{1}{\sqrt{2\Sigma\Delta_r}}\left(-\Delta_r'+\frac{\Delta_r(2r-i\chi)}{\Sigma}\right),\;\; \Delta_{10}=\bar{m}^\mu\partial_\mu+\frac{1}{\sqrt{2\Sigma\Delta_\chi}}\left(\dot{\Delta}_\chi+\frac{5i\Delta_\chi}{r-i\chi}\right),\\
&\Delta_{11}=\bar{m}^\mu\partial_\mu+\frac{1}{\sqrt{2\Sigma\Delta_\chi}}\left(\frac{\dot{\Delta}_\chi}{2}+\frac{ir\Delta_\chi}{\Sigma}\right),\quad\quad\quad\Delta_{12}=n^\mu\partial_\mu+\frac{1}{\sqrt{2\Sigma\Delta_r}}\left(-\frac{\Delta_r'}{2}+\frac{i\chi\Delta_r}{\Sigma}\right),
\end{aligned}
\end{equation}
where primes denote derivatives with respect to $r$ and dots denote derivatives with respect to $\chi$. The differential operators appearing in the Teukolsky equations \eqref{eq:spinPlusTeukolskyeqns} and \eqref{eq:spinMinusTeukolskyeqns} are
\begin{equation}
\begin{aligned}
&\,\mathcal{D}_n=\partial_r+i\frac{K_r(r,\tilde{\omega},k)}{\Delta_r}+n\frac{\Delta_r'}{\Delta_r},\quad\quad\quad\quad \mathcal{D}_n^\dagger=\partial_r-i\frac{K_r(r,\tilde{\omega},k)}{\Delta_r(r)}+n\frac{\Delta_r'}{\Delta_r},\\
&\,\mathcal{L}_n=\partial_{\chi}+\frac{K_\chi(\chi,\tilde{\omega},k)}{\Delta_\chi}+n\frac{\dot{\Delta}_\chi}{\Delta_\chi},\quad\quad\quad\quad \mathcal{D}_n^\dagger=\partial_{\chi}-\frac{K_\chi(\chi,\tilde{\omega},k)}{\Delta_\chi}+n\frac{\dot{\Delta}_\chi}{\Delta_\chi},
\end{aligned}
\end{equation}
where
\begin{equation}
K_r(r,\tilde{\omega},k)=\Xi\left(ka-\tilde{\omega}\left(a^2+r^2\right)\right),\quad\quad\quad K_\chi(\chi,\tilde{\omega},k)=\Xi\left(ka-\tilde{\omega}\left(a^2-\chi^2\right)\right).
\end{equation}

\section{Explicit demonstration of existence of a quasinormal mode} 
\label{app:poleskipping}

\paragraph{}In this Appendix we explain how to obtain the explicit expressions \eqref{eq:exactRminussoln} and \eqref{eq:exactRplussoln} for the solutions to the radial Teukolsky equation at the pole-skipping point $(\tilde{\omega},\lambda)=(\tilde{\omega}_*,\lambda_*)$, and use this to obtain explicit expressions for the quasi-normal mode solutions at and near this point. As mentioned in the main text, our starting point is the observation that the Starobinsky-Teukolsky constant $\mathcal{C}_{ST}^2$ (defined in equation \eqref{eq:STconstantsdefns}) vanishes for $(\tilde{\omega},\lambda)=(\tilde{\omega}_*,\lambda_*)$. As a consequence, we conclude from equation \eqref{eq:eq:STidentity3} that there are solutions $R^{\pm}_{\tilde{\omega}_* k\lambda_*}(r)$ that are annihilated by the 4th order differential operators in the Starobinsky-Teukolsky identities \eqref{eq:STidentity1} and \eqref{eq:STidentity2}:
\begin{equation}
\begin{aligned}
\mathcal{D}_{-1}^\dagger\Delta_r\mathcal{D}_0^\dagger\mathcal{D}_0^\dagger\Delta_r\mathcal{D}_1^\dagger R^+_{\tilde{\omega}_* k \lambda_*}(r)=0,\quad\quad\quad\quad\quad\mathcal{D}_{-1}\Delta_r\mathcal{D}_0\mathcal{D}_0\Delta_r\mathcal{D}_1 R^-_{\tilde{\omega}_* k \lambda_*}(r)=0.
\end{aligned}
\end{equation}

\paragraph{}Each of these is a set of four nested first-order differential equations that can be formally integrated to give general solutions dependent on four integration constants. It is then straightforward to verify by substitution that each of these satisfies the corresponding radial Teukolsky equation for a unique value of these constants (up to an overall normalisation). This procedure yields the solutions parameterised by the constants $\alpha^\pm$ in equations \eqref{eq:exactRminussoln} and \eqref{eq:exactRplussoln}, where the constants $c_n$ are
\begin{equation}
\begin{aligned}
&\,c_3=\frac{1}{3r_0^2}+\frac{12iM\Xi\tilde{\omega}_*^2}{a\lambda_*\left(-k\lambda_*+a\tilde{\omega}_*\left(\lambda_*-6\right)\right)},\quad c_2=\frac{1}{r_0}+\frac{6M\tilde{\omega}_*}{a\left(-k\lambda_*+a\tilde{\omega}_*\left(\lambda_*-6\right)\right)},\\
&\,c_0=\frac{r_0}{3}+\frac{4M\left(-k\lambda_*+a\tilde{\omega}_*\left(\lambda_*-3\right)\right)}{\lambda_*\left(-k\lambda_*+a\tilde{\omega}_*\left(\lambda_*-6\right)\right)}.
\end{aligned}
\end{equation}
An integral expression for the second solution to each Teukolsky equation (those paramaterised by the constants $\beta^\pm$ in equations \eqref{eq:exactRminussoln} and \eqref{eq:exactRplussoln}) can then be identified using the Wronskian method. One can verify that the Starobinsky-Teukolsky identity \eqref{eq:STidentity1} maps the solution parameterised by $\beta^+$ to the solution parameterised by $\alpha^-$ and annihilates the solution parameterised by $\alpha^+$, while the Starobinsky-Teukolsky identity \eqref{eq:STidentity2} maps the solution parameterised by $\beta^-$ to the solution parameterised by $\alpha^+$ and annihilates the solution parameterised by $\alpha^-$.

\paragraph{}As explained in the main text, to identify the quasinormal mode solution at $(\tilde{\omega},\lambda)=(\tilde{\omega}_*,\lambda_*)$, we first impose ingoing boundary conditions at the horizon by setting $\beta^-=0$ in the general solutions \eqref{eq:exactRminussoln} and \eqref{eq:exactRplussoln}. A quasinormal mode then exists for the values of $\alpha^{\pm}$ and $\beta^+$ for which the two conditions \eqref{eq:QNMcond1} and \eqref{eq:QNMcond2} are satisfied. These conditions have the unique solution
\begin{equation}
\begin{aligned}
\label{eq:alphabetasols}
\frac{\beta_+}{\alpha_+}=&\,\Biggr[\frac{a\lambda_*\left(-k\lambda_*+a\omega_*(\lambda_*-6)\right)}{36M\omega_*r_0^2c_3}\exp\left(-2\int^\infty_{r_0}\left(i\frac{K_r(r,\tilde{\omega},k)}{\Delta_r}-\frac{1}{2(r-r_0)}+\frac{1}{2r}\right)dr\right)\\
&\,-\int^\infty_{r_0}\frac{\Delta_r dr}{G(r)^2I(r)^2}\Biggr]^{-1},\\
\frac{\beta_+}{\alpha_-}=&\,\frac{36M\omega_*r_0^2}{c_3a\lambda_*(-k\lambda_*+a\omega_*(\lambda_*-6))}\Biggl[1+\frac{2r_0^4a^2}{M^2L^4}\left(1+\frac{r_0}{iL^2\Xi\omega_*}\right)^2\Bigl\{2\frac{L^2}{r_0^2}-15\left(1+\frac{iL^2\Xi\omega_*}{r_0}\right)\Bigr\}\Biggr],
\end{aligned}
\end{equation}
and thus there is a quasinormal mode at $(\tilde{\omega}_*,\lambda_*)$ for these values of the constants. 

\paragraph{}As described in the main text, we can build on the exact solutions \eqref{eq:exactRminussoln} and \eqref{eq:exactRplussoln} at the pole-skipping point $(\tilde{\omega}_*,\lambda_*)$ to prove that there exist quasi-normal mode solutions perturbatively close to this point. After perturbing $\tilde{\omega}$ and $\lambda$ as in equation \eqref{eq:pertaway}, it is straightforward to identify the ingoing solution for $R^{-}_{\tilde{\omega} k\lambda}(r)$ as
\begin{equation}
\begin{aligned}
R^{-}_{\tilde{\omega} k\lambda}(r)&\,=\alpha^-\frac{(r-r_0)^3}{I(r)\Delta_r(r)}+O(\epsilon),
\end{aligned}
\end{equation}
to leading order in $\epsilon$.

\paragraph{} However, the situation is more subtle for $R^{+}_{\tilde{\omega} k\lambda}(r)$. Far from $r_0$, the corresponding Teukolsky equation can be solved perturbatively to give the general solution \eqref{eq:exactRplussoln} up to $O(\epsilon)$ corrections. However, for any finite $\epsilon$ there is a unique ingoing solution (up to overall normalisation) and we have to therefore identify the unique value of the ratio $\beta^+/\alpha^+$ corresponding to this. To do so, we first expand the general solution \eqref{eq:exactRplussoln} near the horizon, yielding (up to an overall multiplicative constant)
\begin{equation}
\begin{aligned}
\label{eq:smallepssoln}
R^{+}_{\tilde{\omega} k\lambda}\rightarrow (r-r_0)^{-1/2}\Biggl[1+&\,(r-r_0)\Biggl(\frac{\beta_+}{\alpha_+}\frac{4\pi Tr_0(a^2+r_0^2)}{G(r_0)^2}+\frac{G'(r_0)}{G(r_0)}\\
&\,-\frac{4iL^2r_0\Xi\tilde{\omega}_*+3(6r_0^2+L^2+a^2)}{8\pi TL^2(a^2+r_0^2)}\Biggr)+\ldots\Biggr]+O(\epsilon).
\end{aligned}
\end{equation}
We will match this to the near-horizon form of the ingoing solution. To identify this, we first solve the Teukolsky equation near the horizon (for $\epsilon\ne0$) and demand that the corresponding metric perturbations are ingoing. Then, by expanding this ingoing solution at small $\epsilon$ we obtain (up to an overall multiplicative constant)
\begin{equation}
\begin{aligned}
\label{eq:NHingoingsoln}
R^{+}_{\tilde{\omega} k\lambda}\rightarrow &\,(r-r_0)^{-1/2}\Biggl[1+(r-r_0)\Biggl(\frac{\delta\lambda}{\delta\tilde{\omega}}\frac{i}{2\Xi(a^2+r_0^2)}\\
&\,-\frac{(a^2+r_0^2)(4ir_0L^2\Xi\tilde{\omega}_*+3a^2)+3(3a^2(L^2+r_0^2)-\Xi L^2r_0^2)}{8\pi TL^2(a^2+r_0^2)^2}\Biggr)+\ldots\Biggr]+O(\epsilon).
\end{aligned}
\end{equation}
Matching this to \eqref{eq:smallepssoln} and simplifying gives
\begin{equation}
\begin{aligned}
\label{eq:betaratioanswer}
\frac{\beta^+}{\alpha^+}=\frac{G(r_0)^2}{4\pi Tr_0(a^2+r_0^2)}\left[\frac{3r_0}{a^2+r_0^2}-\frac{G'(r_0)}{G(r_0)}+\frac{i}{2\Xi(a^2+r_0^2)}\frac{\delta\lambda}{\delta\tilde{\omega}}\right],
\end{aligned}
\end{equation}
as the ingoing boundary condition. Therefore for any small, non-zero $\epsilon$ the ingoing solution for $R^{+}_{\tilde{\omega} k\lambda}(r)$ is \eqref{eq:exactRplussoln}, with $\beta_+/\alpha_+$ given by \eqref{eq:betaratioanswer}.

\paragraph{}Having identified the unique ingoing solutions for $R^{\pm}_{\tilde{\omega} k\lambda}(r)$ to leading order in $\epsilon$, we can now prove that there must exist quasinormal modes whose dispersion relations pass through the pole-skipping point $(\tilde{\omega}_*,\lambda_*)$. To do so, we simply take the ingoing solution we have just constructed and expand it near the asymptotic boundary. A quasinormal mode exists when the two conditions \eqref{eq:QNMcond1} and \eqref{eq:QNMcond2} are satisfied, which we find is the case when
\begin{equation}
\begin{aligned}
\label{eq:QNMslope}
\frac{\delta{\lambda}}{\delta\tilde{\omega}}&\,=-2i\Xi(a^2+r_0^2)\Biggl\{\frac{G'(r_0)}{G(r_0)}-\frac{3r_0}{a^2+r_0^2}+\frac{4\pi Tr_0(a^2+r_0^2)}{G(r_0)^2}\Biggl[-\int^\infty_{r_0}\frac{\Delta_r dr}{G(r)^2I(r)^2}\\
&\,+\frac{a\lambda_*\left(-k\lambda_*+a\omega_*(\lambda_*-6)\right)}{36M\omega_*r_0^2c_3}\exp\left(-2\int^\infty_{r_0}\left(i\frac{K_r(r,\tilde{\omega},k)}{\Delta_r}-\frac{1}{2(r-r_0)}+\frac{1}{2r}\right)dr\right)\Biggr]^{-1}\Biggr\}.
\end{aligned}
\end{equation}
In other words, there must be a quasinormal mode dispersion relation $\tilde{\omega}(k,\lambda)$ passing through the point $(\tilde{\omega}_*,\lambda_*)$, with slope given by equation \eqref{eq:QNMslope}. More generally, the dependence of the ingoing solution upon the slope $\delta\lambda/\delta\tilde{\omega}$ means that we can construct ingoing solutions near $(\tilde{\omega}_*,\lambda_*)$ satisfying any asymptotic boundary condition, not just that corresponding to a quasi-normal mode, by choosing an appropriate value of the slope $\delta\lambda/\delta\tilde{\omega}$.

\bibliographystyle{JHEP}
\bibliography{RotatingBTZ}

 \end{document}